\input harvmac.tex

\input epsf.tex

\def\figin{\epsfcheck\figin}\def\figins{\epsfcheck\figins}
\def\epsfcheck{\ifx\epsfbox\UnDeFiNeD
\message{(NO epsf.tex, FIGURES WILL BE IGNORED)}
\gdef\figin##1{\vskip2in}\gdef\figins##1{\hskip.5in}
\else\message{(FIGURES WILL BE INCLUDED)}%
\gdef\figin##1{##1}\gdef\figins##1{##1}\fi}
\def\DefWarn#1{}
\def\figinsert{\goodbreak\midinsert}
\def\ifig#1#2#3{\DefWarn#1\xdef#1{figure~\the\figno}
\writedef{#1\leftbracket figure\noexpand~\the\figno}%
\figinsert\figin{\centerline{#3}}\medskip\centerline{\vbox{\baselineskip12pt
\advance\hsize by -1truein\noindent\footnotefont{\bf
Figure~\the\figno:} #2}}
\bigskip\endinsert\global\advance\figno by1}


\def\pibar{{\bar\pi}}
\def\xb{{\bar\xi}}
\def\psibar{{\bar\psi}}

\def\t{{\theta}}
\def\tto{{\tilde\theta^1}}
\def\tb{{\bar\theta}}
\def\th{{\hat\theta}}
\def\a{\alpha}
\def\k{\kappa}

\def\ah{{\hat\alpha}}
\def\mh{{\hat m}}
\def\nh{{\hat n}}
\def\ad{{\dot a}}
\def\adot{{\dot a}}
\def\bd{{\dot b}}
\def\b{\beta}
\def\s{\sigma}
\def\g{\gamma}
\def\G{\Gamma}
\def\d{\delta}
\def\e{\epsilon}
\def\ve{\varepsilon}
\def\l{\lambda}
\def\lh{{\hat\lambda}}
\def\dh{{\hat\delta}}
\def\gh{{\hat\gamma}}
\def\eh{{\hat\epsilon}}
\def\bh{{\hat\beta}}
\def\gh{{\hat\gamma}}
\def\mt{{\tilde {\mu}}}
\def\xt{{\tilde x}}
\def\tt{{\tilde \theta}}
\def\p{\partial}
\def\pb{\bar\partial}
\def\n{\nabla}

\def\half{{1 \over 2 }}


\lref\HullVG{
  C.~M.~Hull,
 ``Timelike T-duality, de Sitter Space, Large N Gauge Theories and
  Topological Field Theory,''
  JHEP {\bf 9807}, 021 (1998)
  arXiv:hep-th/9806146.
}

\lref\MetsaevIT{
  R.~R.~Metsaev and A.~A.~Tseytlin,
  ``Type IIB Superstring Action in AdS(5) x S(5) Background,''
  Nucl.\ Phys.\  B {\bf 533}, 109 (1998)
  arXiv:hep-th/9805028.
}
\lref\bershadsky{
  N.~Berkovits, M.~Bershadsky, T.~Hauer, S.~Zhukov and B.~Zwiebach,
  ``Superstring Theory on AdS(2) x S(2) as a Coset Supermanifold,''
  Nucl.\ Phys.\  B {\bf 567}, 61 (2000)
  hep-th/9907200.
}

\lref\fms{D. Friedan, S. H. Shenker , E. J. Martinec,
`` Covariant Quantization of Superstrings'',
 Phys.Lett.B160:55,1985.}

\lref\ademollo{  M. Ademollo et al,
``Dual String with U(1) Color Symmetry'',
Nucl. Phys. B111 (1976) 77.}

\lref\oogv{ H. Ooguri and C. Vafa, ``Geometry of N=2 Strings'',
Nucl. Phys. B361 (1991) 469.}

\lref\symtwistor{
N. Berkovits, ``Self-Dual Super-Yang-Mills as a String Theory'', JHEP 0405 (2004) 034,
 hep-th/0403280}

\lref\supersieg{
W. Siegel, ``Green-Schwarz Formulation of Self-Dual Superstring'',Phys. Rev. D47 (1993) 2512,
  hep-th/9210008}

\lref\vafatopo{
N. Berkovits, C. Vafa, ``N=4 Topological Strings'', Nucl. Phys. B433 (1995) 123,
  hep-th/9407190}

\lref\seiberg{
N. Berkovits, N. Seiberg, ``Superstrings in graviphoton background and N=1/2 + 3/2 supersymmetry'',
  JHEP 0307:010,2003.
  hep-th/0306226}

\lref\siegelproj{ M. Hatsuda and W. Siegel,``A New
Holographic Limit of $AdS_5\times S^5$'', Phys. Rev. D67
(2003) 066005, hep-th/0211184.}

\lref\howeproj{ P. Heslop and P.S. Howe, ``Chiral Superfields
in IIB Supergravity'', Phys. Lett. B502 (2001) 259, hep-th/0008047.}

\lref\seibergone{
N. Seiberg, `` Noncommutative superspace, N = 1/2 supersymmetry, field theory and string theory'',
  JHEP 0306:010,2003, hep-th/0305248}

\lref\oog{
H. Ooguri and  Cumrun Vafa, ``The C deformation of Gluino and nonplanar diagrams'',
  Adv.Theor.Math.Phys.7:53-85,2003, hep-th/0302109}

\lref\wessbagger{
J. Wess and  J. Bagger, ``Supersymmetry and supergravity'',
Princeton, USA: Univ. Pr. (1992) 259 p.}

\lref\aldaylectures{
  L.~F.~Alday,
  ``Lectures on Scattering Amplitudes via AdS/CFT'',
  arXiv:0804.0951 [hep-th].
}

\lref\stieberger{
S. Stieberger and T. R. Taylor, ``Supersymmetry Relations and MHV Amplitudes in Superstring Theory'',
 Nucl.Phys.B793:83-113,2008,
  arXiv:0708.0574 [hep-th]}

\lref\verlinde{
E. Verlinde, H. L. Verlinde, ``Lectures On String Perturbation Theory'',
  Trieste School 1988: Superstrings:189 (QCD161:T7322:1988)}

\lref\jmjr{
  J.~M.~Maldacena and J.~G.~Russo,
  ``Large N limit of non-commutative gauge theories,''
  JHEP {\bf 9909}, 025 (1999)
  arXiv:hep-th/9908134.
}

\lref\mcgreevy{
  J.~McGreevy and A.~Sever,
  ``Planar scattering amplitudes from Wilson loops,''
  arXiv:0806.0668 [hep-th].
}
\lref\aldayjmtwo{
  L.~F.~Alday and J.~M.~Maldacena,
  ``Comments on operators with large spin,''
  JHEP {\bf 0711}, 019 (2007)
  arXiv:0708.0672 [hep-th].
}

\lref\jmcn{
  J.~M.~Maldacena and C.~Nunez,
  ``Supergravity description of field theories on curved manifolds and a no  go
  theorem,''
  Int.\ J.\ Mod.\ Phys.\  A {\bf 16}, 822 (2001)
  arXiv:hep-th/0007018.
}
\lref\hlat{
  H.~Liu and A.~A.~Tseytlin,
  ``D3-brane D-instanton configuration and N = 4 super YM theory in  constant
  self-dual background,''
  Nucl.\ Phys.\  B {\bf 553}, 231 (1999)
  arXiv:hep-th/9903091.
}
\lref\policastro{
  R.~Benichou, G.~Policastro and J.~Troost,
  ``T-duality in Ramond-Ramond backgrounds,''
  arXiv:hep-th/0801.1785.
}
\lref\SchwarzTE{
  A.~S.~Schwarz and A.~A.~Tseytlin,
  ``Dilaton shift under duality and torsion of elliptic complex,''
  Nucl.\ Phys.\  B {\bf 399}, 691 (1993)
  arXiv:hep-th/9210015.
}

\lref\tseytlintdual{
  R.~Ricci, A.~A.~Tseytlin and M.~Wolf,
  ``On T-Duality and Integrability for Strings on AdS Backgrounds,''
  JHEP {\bf 0712}, 082 (2007)
  arXiv:0711.0707 [hep-th].
}

\lref\polch{   I.~Bena, J.~Polchinski and R.~Roiban,
  ``Hidden symmetries of the AdS(5) x S**5 superstring,''
  Phys.\ Rev.\  D {\bf 69}, 046002 (2004)
  arXiv:hep-th/0305116.
}

\lref\KruczenskiFB{
  M.~Kruczenski,
  ``A note on twist two operators in N = 4 SYM and Wilson loops in Minkowski
  signature,''
  JHEP {\bf 0212}, 024 (2002)
  arXiv:hep-th/0210115.
}
\lref\fajm{  L.~F.~Alday and J.~Maldacena,
  ``Comments on gluon scattering amplitudes via AdS/CFT,''
  JHEP {\bf 0711}, 068 (2007)
  arXiv:0710.1060 [hep-th].
}

\lref\brandhuber{
  A.~Brandhuber, P.~Heslop and G.~Travaglini,
  ``MHV Amplitudes in N=4 Super Yang-Mills and Wilson Loops,''
  Nucl.\ Phys.\  B {\bf 794}, 231 (2008)
  arXiv:0707.1153 [hep-th].
}

\lref\FerberQX{
  A.~Ferber,
  ``Supertwistors And Conformal Supersymmetry,''
  Nucl.\ Phys.\  B {\bf 132}, 55 (1978).
}

\lref\abdk{
  C.~Anastasiou, Z.~Bern, L.~J.~Dixon and D.~A.~Kosower,
  ``Planar amplitudes in maximally supersymmetric Yang-Mills theory,''
  Phys.\ Rev.\ Lett.\  {\bf 91}, 251602 (2003)
  arXiv:hep-th/0309040.
}
\lref\bds{
  Z.~Bern, L.~J.~Dixon and V.~A.~Smirnov,
  ``Iteration of planar amplitudes in maximally supersymmetric Yang-Mills
  theory at three loops and beyond,''
  Phys.\ Rev.\  D {\bf 72}, 085001 (2005)
  arXiv:hep-th/0505205.
}

\lref\Nair{
  V.~P.~Nair,
  ``A current algebra for some gauge theory amplitudes'',
  Phys.\ Lett.\  B {\bf 214}, 215 (1988).
}

\lref\polyakov{
 A.~M.~Polyakov,
  ``The wall of the cave,''
  Int.\ J.\ Mod.\ Phys.\  A {\bf 14}, 645 (1999)
  arXiv:hep-th/9809057.
  }

\lref\sokatch{
  J.~M.~Drummond, J.~Henn, V.~A.~Smirnov and E.~Sokatchev,
  ``Magic identities for conformal four-point integrals,''
  JHEP {\bf 0701}, 064 (2007)
  arXiv:hep-th/0607160.
}

\lref\nmhv{
  J.~M.~Drummond, J.~Henn, G.~P.~Korchemsky and E.~Sokatchev,
  ``Dual superconformal symmetry of scattering amplitudes in N=4
  super-Yang-Mills theory,''
  arXiv:0807.1095 [hep-th].
}
\lref\BuscherQJ{
  T.~H.~Buscher,
``Path Integral Derivation of Quantum Duality in Nonlinear Sigma Models,''
  Phys.\ Lett.\  B {\bf 201}, 466 (1988).
}

\lref\sixwilson{
  J.~M.~Drummond, J.~Henn, G.~P.~Korchemsky and E.~Sokatchev,
  ``Hexagon Wilson loop = six-gluon MHV amplitude,''
  arXiv:0803.1466 [hep-th].
}

\lref\sixgluon{
  Z.~Bern, L.~J.~Dixon, D.~A.~Kosower, R.~Roiban, M.~Spradlin, C.~Vergu and A.~Volovich,
  ``The Two-Loop Six-Gluon MHV Amplitude in Maximally Supersymmetric Yang-Mills
  Theory,''
  arXiv:0803.1465 [hep-th].
}
\lref\Cremmer{
  E.~Cremmer and B.~Julia,
  ``The N=8 Supergravity Theory. 1. The Lagrangian,''
  Phys.\ Lett.\  B {\bf 80}, 48 (1978).
  Nucl.\ Phys.\  B {\bf 159}, 141 (1979).
}

\lref\MandalFS{
  G.~Mandal, N.~V.~Suryanarayana and S.~R.~Wadia,
  ``Aspects of semiclassical strings in AdS(5),''
  Phys.\ Lett.\  B {\bf 543}, 81 (2002)
  arXiv:hep-th/0206103.
}

\lref\DrummondAUA{
  J.~M.~Drummond, G.~P.~Korchemsky and E.~Sokatchev,
  ``Conformal properties of four-gluon planar amplitudes and Wilson loops,''
  Nucl.\ Phys.\  B {\bf 795}, 385 (2008)
  arXiv:0707.0243 [hep-th].
}

\lref\DrummondCF{
  J.~M.~Drummond, J.~Henn, G.~P.~Korchemsky and E.~Sokatchev,
  ``On planar gluon amplitudes/Wilson loops duality,''
  Nucl.\ Phys.\  B {\bf 795}, 52 (2008)
  arXiv:0709.2368 [hep-th].
}

\lref\sokatchevward{
  J.~M.~Drummond, J.~Henn, G.~P.~Korchemsky and E.~Sokatchev,
  ``Conformal Ward identities for Wilson loops and a test of the duality with
  gluon amplitudes,''
  arXiv:0712.1223 [hep-th].
}

\lref\farr{
  L.~F.~Alday and R.~Roiban,
  ``Scattering Amplitudes, Wilson Loops and the String/ Gauge Theory
  Correspondence,''
  arXiv:0807.1889 [hep-th].
}

\lref\BernQK{
  Z.~Bern, G.~Chalmers, L.~J.~Dixon and D.~A.~Kosower,
  ``One loop N gluon amplitudes with maximal helicity violation via collinear
  limits,''
  Phys.\ Rev.\ Lett.\  {\bf 72}, 2134 (1994)
  arXiv:hep-ph/9312333.
}

\lref\TseytlinNew{
N. Beisert, R. Ricci, A. A. Tseytlin and M. Wolf,
`` World sheet dualities and integrability of the $AdS_5 \times S^5$ superstring
 sigma model,''
arXiv:0807.3228[hep-th]
}

\lref\Howewest{
P. Howe and P. West,
``The complete N=2 D=10 supergravity,''
Nucl. Phys. B238, 181 (1984).
}

\lref\ParkeGB{
  S.~J.~Parke and T.~R.~Taylor,
  ``An Amplitude for $n$ Gluon Scattering,''
  Phys.\ Rev.\ Lett.\  {\bf 56}, 2459 (1986).
}

\lref\AganagicYH{
  M.~Aganagic and C.~Vafa,
  ``Mirror Symmetry and Supermanifolds,''
  arXiv:hep-th/0403192.
}

\lref\howeb{ N. Berkovits and P.S. Howe,
``Ten-Dimensional Supergravity
Constraints from the Pure Spinor Formalism for the Superstring,''
Nucl. Phys. B635 (2002) 75, arXiv:hep-th/0112160.}

\lref\BCOV{
  M.~Bershadsky, S.~Cecotti, H.~Ooguri and C.~Vafa,
  ``Holomorphic Anomalies in Topological Field Theories,''
  Nucl.\ Phys.\  B {\bf 405}, 279 (1993)
  arXiv:hep-th/9302103\semi
  I.~Antoniadis, E.~Gava, K.~S.~Narain and T.~R.~Taylor,
  ``Topological Amplitudes in String Theory,''
  Nucl.\ Phys.\  B {\bf 413}, 162 (1994)
  arXiv:hep-th/9307158.
}

\lref\oda{
I. Oda and M. Tonin,
``On the Berkovits Covariant Quantization of GS Superstring,''
Phys. Lett. B520 (2001) 398,
arXiv:hep-th/0109051.}

\lref\lads{
N. Berkovits,
``Quantum Consistency of the Superstring in $AdS_5\times S^5$
Background,''
JHEP 0503 (2005) 041,
arXiv:hep-th/0411170.}

\lref\cangemi{
  D. Cangemi,
``Self-Dual Yang-Mills Theory and One-Loop Maximally Helicity Violating
Multi-Gluon Amplitudes,''
  Nucl.\ Phys.\  B {\bf 484}, 521 (1997),
  arXiv:hep-th/9605208.
}

\lref\chalmers{
  G. Chalmers and W. Siegel,
``The Self-Dual Sector of QCD
Amplitudes,''
  Phys.\ Rev.\  D {\bf 54}, 7628 (1996),
  arXiv:hep-th/9606061.
}

\lref\bern{
 Z. Bern, L. Dixon, D. Dunbar and D. Kosower,
``One-Loop Self-Dual and N=4 Super-Yang-Mills,''
  Phys.\ Lett.\  B {\bf 394}, 105 (1997),
  arXiv:hep-th/9611127.
}

\lref\BergshoeffPV{
  E.~Bergshoeff, R.~Kallosh, T.~Ortin, D.~Roest and A.~Van Proeyen,
  ``New formulations of D = 10 supersymmetry and D8 - O8 domain walls,''
  Class.\ Quant.\ Grav.\  {\bf 18}, 3359 (2001)
  arXiv:hep-th/0103233.
}

\lref\damour{
  T.~Damour, A.~Kleinschmidt and H.~Nicolai,
  ``Constraints and the E10 Coset Model,''
  Class.\ Quant.\ Grav.\  {\bf 24}, 6097 (2007)
  arXiv:0709.2691 [hep-th].
}

\lref\west{
  F.~Riccioni and P.~West,
  ``The E(11) origin of all maximal supergravities,''
  JHEP {\bf 0707}, 063 (2007)
  arXiv:0705.0752 [hep-th].
}




\Title{\vbox{\baselineskip12pt\hbox{} \hbox{IFT-P.015.2008
}}} {\vbox{ \vskip -5cm {\centerline{ Fermionic T-Duality,}
\centerline{  Dual Superconformal Symmetry, and
}  \centerline{   the Amplitude/Wilson Loop Connection }
 }}}

\vskip  -5mm
 \centerline{ Nathan Berkovits$^{a}$ and Juan Maldacena$^{b}$}

\bigskip
{\sl \centerline{$^a$Instituto de F\'{\i}sica Te\'orica,
State University of S\~ao Paulo,
S\~ao Paulo, SP 01405, BRASIL  } \centerline{
 nberkovi@ift.unesp.br }
\medskip
\centerline{$^{b}$School of Natural Sciences, Institute for
Advanced Study, Princeton, NJ 08540, USA} \centerline{
  malda@ias.edu}
  }


\leftskip 8mm  \rightskip 8mm \baselineskip14pt \noindent
%
 We show that tree level superstring theories on certain supersymmetric backgrounds
 admit a symmetry which we call ``fermionic T-duality''. This is a non-local
 redefinition of the fermionic worldsheet fields similar to the redefinition we perform
 on bosonic variables when we do an ordinary T-duality. This duality maps a supersymmetric
 background to another supersymmetric background with different RR fields and a different dilaton.
 We show that a certain combination of bosonic and fermionic T-dualities
 maps the full superstring theory on $AdS_5 \times S^5$ back to itself in such a way that
  gluon scattering
 amplitudes in the original theory  map to something very close to Wilson loops in the dual
 theory. This duality maps the ``dual superconformal symmetry'' of the original theory
 to the ordinary superconformal symmetry of the dual model. This explains the dual
 superconformal invariance of planar scattering amplitudes of $N=4$ super Yang Mills and
also sheds some light on
the connection between amplitudes and Wilson loops.
In the appendix, we propose a simple prescription for open superstring
MHV tree amplitudes in a flat background.

\bigskip\medskip

\leftskip 0mm  \rightskip 0mm
 \Date{\hskip 8mm July 2008}

\newsec{Introduction}

During the past year a surprising connection was found between
planar scattering amplitudes and Wilson loops in ${\cal N}=4$
super Yang Mills, for a recent review and a more complete set of references
see \farr . This was first noticed in the strong coupling
computation of the amplitudes in \fajm . The connection that was
found in \fajm\ was apparently
 valid only at leading
order in the strong coupling expansion. However, the same
connection was soon found in weak coupling computations
\refs{\DrummondAUA,\brandhuber,\DrummondCF,\sokatchevward}, based on previous
amplitude computations in \refs{\abdk,\bds}.  More recently an
impressive check of this relationship was performed at two loops
for six gluons in \refs{\sixgluon,\sixwilson}.

\ifig\wilsonloop{ Relation between the amplitude and the Wilson loop. A planar
scattering amplitude of $n$ gluons is related to a  Wilson loop computation involving
an $n$ sided polygonal Wilson loop where the sides are light like vectors given by the
momenta.
  } {\epsfxsize2.5in\epsfbox{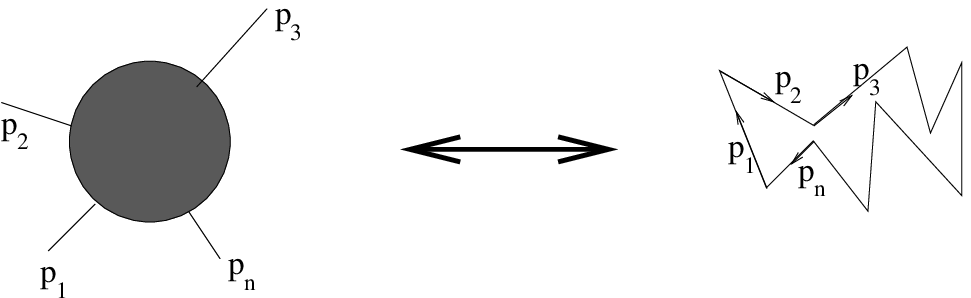}}

The basic statement  of the relationship is as follows. One
considers the color ordered  amplitudes ${\cal A}(P_1, \cdots ,
p_n)$, defined via a color decomposition of the planar amplitude
\eqn\amp{ {\cal A}(a_1,p_1, a_2,p_2 , \cdots) = \sum_{Permutations}
Tr[T^{a_1} \cdots T^{a_n} ] {\cal A}(p_1, \cdots , p_n) } where $a_i$ are
the group indices and $p_i$ are the momenta and we suppressed the
polarization dependence. We can then write the MHV amplitudes as
\eqn\mhvamp{ {\cal A}_{MHV} = { \cal A}_{MHV,\, tree} \, \, \hat
{\cal A} (p_1,\cdots , p_n) } where ${\cal A}_{MHV, \, tree}$ is the tree level
MHV amplitude \ParkeGB . Then the observation is that
\eqn\relampl{
 \hat { \cal A} (p_1 , \cdots , p_n) =  \langle W(p_1, \cdots , p_n) \rangle
 }
 where $W$ is a Wilson loop that ends on a contour made by $n$ lightlike segments, each
 proportional to $p_i$, see \wilsonloop . To be more precise, the left hand
 side in \relampl\ is infrared divergent and the right hand side is UV divergent. The
 structure of these divergencies is known. The statement is really
  about the finite parts of the
 amplitudes, which can have a complicated dependence on the kinematic invariants of the process.

A closely related   fact is that scattering amplitudes display an interesting
non-trivial symmetry called ``dual conformal invariance''. This symmetry was first
found in perturbative computations in  \sokatch ,
 and it was recently also observed in next to MHV amplitudes in \nmhv , where it was
 promoted to a full ``dual superconformal symmetry''. One would also like to understand
 the origin of this symmetry. If one accepts the relationship between amplitudes and Wilson loops,
 then this symmetry is the ordinary superconformal symmetry acting on Wilson loops.

In this paper we show that one can
understand  this ``dual superconformal symmetry''  using  a
T-duality symmetry of the full superstring   theory on $AdS_5
\times S^5$. The T-duality involves  ordinary bosonic T-dualities,
which were considered already in \fajm , plus   novel
``fermionic'' T-dualities. These fermionic T-dualities consist in
certain non-local redefinitions of the fermionic variables of the
superstring. The fermionic T-dualities change the dilaton and the RR fields without
modifying the metric.  After these T-dualities the sigma model looks the
same as the original sigma model but the computation of the
amplitude in the original model maps to  a computation of an object very close
to a Wilson loop in the dual theory. The ordinary superconformal invariance of the
T-dual model is the ``dual superconformal symmetry'' of the original theory.
The amplitude computation does not map precisely to a Wilson loop computation but
to a certain computation involving D(-1) branes and strings stretching between them.
For MHV amplitudes we expect, on the basis of perturbative computations
\refs{\DrummondAUA,\brandhuber,\sixgluon,\sixwilson},  that the
difference should amount to a simple prefactor which is equal to the tree level MHV
amplitude \mhvamp . We will not derive this factor in this paper,
 but we will give some plausibility
arguments.

We will discuss in some detail the nature of   fermionic T-duality for general
backgrounds.
We will give some general rules regarding the transformation of
the background fields under fermionic T-dualities. Fermionic
T-dualities are possible when we have a supercharge that
anticommutes to zero, $Q^2 =0$. In that case one can represent the
action of this supercharge as the shift of a certain fermionic
coordinate of the sigma model $\theta \to \theta + \epsilon$. The
fermionic T-duality is a transformation of the fermionic variables
rather similar to the one we do for the case of bosonic
T-dualities, in the sense that we redefine the field in such a way
that we exchange the equation of motion with the Bianchi identity.
The T-dual sigma model   leads to a different background for the
superstring. Thus a fermionic T-duality   relates the
superstring on a  supersymmetric background to the
superstring on another supersymmetric background. In general
this fermionic T-duality is a valid symmetry only at string tree
level for reasons similar to the ones that imply that a bosonic
T-duality of a non-compact scalar is only a symmetry of certain tree level
computations.

 A connection between amplitudes and Wilson loops in momentum space was discussed in
 \refs{\polyakov,\mcgreevy}. One performs a Fourier transformation of an ordinary
 Wilson loop to obtain the Wilson loop in momentum space. Then the amplitude is related
 to a particular momentum space Wilson loop which
   looks exactly as the
 one in  \wilsonloop .  The important point we are making here is
 that
 this {\it momentum } space Wilson loop can be computed by mapping it to an ordinary {\it position }
 space Wilson loop with the same shape.

 It was also expected  that ``dual conformal symmetry'' should be connected to integrability.
 In fact, in the simpler case of a bosonic $AdS$ sigma model we show
 that the non-trivial generators in the ``dual conformal group'' correspond to
 some of the non-local
 charges that arise due to integrability. This  conclusion has also been reached
 for the full $AdS_5 \times S^5$ theory in \TseytlinNew .

 In an appendix we also propose a simple
prescription for computing MHV tree level open string
 scattering amplitudes
 in flat space.
 This prescription is related to the self-dual N=2 string
\refs{\ademollo,\oogv} and reproduces
the amplitudes that have been computed previously
 in \stieberger\ using the standard
 formalism.

This paper is organized as follows. In section two we introduce the concept of a
 ``fermionic T-duality'' and we explore some of its properties.
 In section three we perform a set of bosonic and fermionic transformations that
 map $AdS_5 \times S^5$ back to itself which maps the problem of amplitudes to a
 problem closely related to Wilson loops.
 In section four we discuss in more detail what the computation of the amplitude
 maps into.
 In section five we discuss the relation between the conformal symmetry of the dual
 theory and the non-local charges associated to integrability.
 In section six we present some conclusions.

 In Appendix A we discuss a proposal for computing MHV amplitudes in flat space string theory.
 This is  disconnected from the rest of the paper and can be read on its own.

\newsec{Fermionic T-duality}

In this section, we discuss ``fermionic T-duality'' which
is a generalization of the Buscher version of T-duality
to theories with fermionic worldsheet scalars \foot{  T-dualities involving fermionic
fields were considered in \AganagicYH ,  but in their case they were T-dualizing the
phase of a fermionic field, which was essentially bosonic. Thus it is not obviously related to
what we are doing here.  }.  We first show
how fermionic T-duality transforms background superfields
in the Green-Schwarz and pure spinor
sigma models. We then translate these transformations into
the language of Type II supergravity fields and show
that fermionic T-duality changes the background values of the dilaton
and Ramond-Ramond fields. A simple example
of fermionic T-duality relates the flat background
with the self-dual graviphoton background.
In the following section we will apply this transformation to the $AdS_5 \times S^5$ case.

\subsec{Review of bosonic T-duality}

In the sigma model description of T-duality, one starts with a
sigma model \eqn\bos{S = \int d^2 z (g_{mn}(x) + b_{mn}(x))\p x^m
\bar\p x^n} and assumes that the background fields $g_{mn}$ and
$b_{mn}$ are invariant under the shift isometry \eqn\shiftb{x^1\to
x^1 +c, \quad x^\mh \to x^\mh} where $c$ is a constant and $\mh$
ranges over all values except $m=1$. Since $x^1$ only appears with
derivatives, the action is \eqn\bostwo{S = \int d^2 z (g_{11}(\hat
x)\p x^1 \bar\p x^1 + l_{1\mh}(\hat x)\p x^1 \bar\p x^\mh + l_{\mh
1}(\hat x)\p x^\mh \bar\p x^1 + l_{\mh \hat n}(\hat x)\p x^\mh
\bar\p x^\nh)} where $l_{mn} = g_{mn} + b_{mn}$.

If $g_{11}$ is nonzero, one can use the Buscher procedure \BuscherQJ\ to
T-dualize the sigma model with respect to $x^1$.
This is done as follows. We first replace the derivatives of $x^1$ by a vector field $(A, \bar A)$
on
the worldsheet and we add a lagrange multiplier field  $\widetilde x^1 $, that forces the vector field
to be the derivative of a scalar
\eqn\actionbg{S = \int d^2 z [g_{11} A \bar A  +
l_{1\hat m} A \pb x^\mh + l_{\mh 1}\p x^\mh \bar A + l_{\mh \nh}\p x^\mh
\pb x^\nh
+ \widetilde x^1 (\p \bar A - \bar\p A)]}
If we first integrate out the lagrange multiplier $\widetilde
x^1$, we force $\p \bar A - \bar \p A =0$ which can be solved by
saying that $A = \partial x^1 $ and $\bar A = \bar \p x^1$ and we
go back to the original model. On the other hand, if we first
integrate out the vector field we obtain the T-dualized action
\eqn\actionbgg{S = \int d^2 z [g'_{11} \p \widetilde x^1 \pb
\widetilde x^1 + l'_{1\mh} \p \widetilde x^1\pb x^\mh + l'_{\mh
1}\p x^\mh\pb  \widetilde x^1 + l'_{\mh\nh}\p x^\mh \pb x^\nh]}
where \eqn\primefieldsb{g'_{11} = ( g_{11})^{-1}, \quad l'_{1\mh}
=  (g_{11})^{-1} l_{1\mh}, \quad l'_{\mh 1} = - (g_{11})^{-1}
l_{\mh 1}, }
$$
l'_{\mh\nh} = l_{\mh\nh} - (g_{11})^{-1}
l_{\mh 1} l_{1\nh}.$$
Furthermore, the measure factor coming from integration over the bosonic
vector field will induce a change in the dilaton $\phi$ by \refs{\BuscherQJ,\SchwarzTE}
\eqn\dilatonb{\phi' = \phi - \half \log g_{11}.}

In the above discussion we have not said whether $x^1$ is compact or not. In order
for the transformation to be valid on an arbitrary compact Riemman surface, it is
important that $x^1$ is compact. The reason is that on an arbitrary surface,
the condition that the field strength of the vector field is zero does not
imply that
it is the gradient of a scalar. The vector field could have holonomies on the various
cycles of the Riemann surface. If the Lagrange multiplier field $\widetilde x^1$ is a compact
field that can have winding on these circles, then we   find that, after integrating it out,
it  imposes that the holonomy of
the vector field has  certain integral values. In this case
we can still write the vector field in terms of a scalar $x^1$, which might
wind along the cycles of the Riemman surface.

If we are considering the theory on the sphere or the disk, we do not need to worry about this
and we can perform this transformation even for non-compact scalars, as long as the external
vertex operators do not carry momentum. Note that in this case we can always write a vector field with zero field strength in terms
of the gradient of a scalar. If we are on the sphere and the external vertex
operators carry momentum, then the T-dual problem does not correspond to anything we ordinarily
encounter in string theory. The situation is nicer in the case of the disk with external states
that carry momentum only at the boundary of the disk. In this case, after the T-duality these
 open string states carry winding and we can interpret them as stretching between different D-branes that
are localized in the T-dual coordinate. In general we will get as many D-branes as insertions
we have on the boundary.
  In this case  we need
to treat the zero modes of the scalars separately. The original model contains an integration
of the zero mode of the scalars which needs to be done before doing the T-duality. Correspondingly
in the T-dual model we do not integrate over the zero mode of the T-dualized scalar, we just
set it to some arbitrary value at some point on the boundary of the disk. This fixes the
position of one of the D-branes on the T-dual circle. The other D-brane positions are fixed
by the momenta that the vertex operators carried in the original theory.

In summary, even though a bosonic T-duality for a non-compact scalar is not well defined to
all orders in string perturbation theory, one can do it for the disk diagram (and also for
the sphere if none of the particles carries momentum in the original direction).



\subsec{Sigma model in superspace and fermionic T-duality}

Suppose one is now given a Green-Schwarz-like sigma model depending
on bosonic and fermionic worldsheet variables $(x^m,\t^\mu)$ such
that the worldsheet action is invariant under a constant shift of
one of the fermionic variables $\t^1$. In other words, the
action is invariant under
\eqn\shift{\t^1\to \t^1 +\rho, \quad x^m \to x^m, \quad \t^\mt \to \t^\mt}
where $\rho$ is a fermionic constant and $\mt$ ranges over all fermionic
variables except for $\t^1$. Of course such backgrounds preserve a supersymmetry,
whose properties we will discuss in more detail below.

Invariance under \shift\
implies that $\t^1$ only appears in the action with derivatives
as $\p\t^1$ or $\pb\t^1$, so the worldsheet action has the form
\eqn\action{S = \int d^2 z [B_{11}(Y) \p\t^1\pb\t^1 +
L_{1M}(Y) \p\t^1\pb Y^M + L_{M1}(Y)\p Y^M\pb \t^1 + L_{MN}\p Y^M \pb Y^N]}
where $Y^M=(x^m,\t^\mt)$, $M=(m,\mt)$ ranges over all indices except
for $\mu=1$,
and $L_{MN}(Y)=G_{MN}(Y)+B_{MN}(Y)$ is the sum of the
graded-symmetric tensor $G_{MN}$ and the graded-antisymmetric tensor
$B_{MN}$.

If $B_{11}(Y)$ is nonzero, one can use the Buscher procedure to
T-dualize the sigma model with respect to $\t^1$. This is done by
first introducing a fermionic vector field $(A,\bar A)$. We replace
the derivatives of $\theta^1$ by the fermionic vector field. In addition
we introduce the lagrange multiplier field $\widetilde \theta^1$ which imposes
that the vector field is the derivative of a fermionic scalar via a term
$\int d^2 z \widetilde \theta^1 (\p \bar A - \bar\p A)$.
 The resulting action is
\eqn\actiong{\eqalign{ S = & \int d^2 z [B_{11}(Y) A \bar A  +
L_{1M}(Y)A \pb Y^M + L_{M1}(Y)\p Y^M \bar A + L_{MN}\p Y^M \pb Y^N
\cr
 & ~~~~~~~~ + \tilde \theta^1 (\p \bar A - \bar\p A)]
 }}
 Integrating out $\tilde \theta^1$ imposes that $A = \partial \theta^1$ and $\bar A = \bar \partial
 \theta^1 $.
On the other hand, when
we  first integrate  out the fermionic gauge field
 we obtain the T-dualized action
\eqn\actiongg{S = \int d^2 z [B'_{11}(Y) \p \widetilde \theta^1 \pb\widetilde \theta^1 +
L'_{1M}(Y) \p\widetilde \theta^1 \pb Y^M + L'_{M1}(Y)\p Y^M\pb
\widetilde \theta^1 + L'_{MN}\p Y^M \pb Y^N]}
where
\eqn\primefields{B'_{11} = - (B_{11})^{-1}, \quad
L'_{1M} =  (B_{11})^{-1} L_{1M}, \quad
L'_{M1} =  (B_{11})^{-1} L_{M1}, }
$$
L'_{MN} = L_{MN}
- { 1 \over B_{11} } L_{1 N} L_{M 1 }
$$
Furthermore,
the measure factor coming from integration over the fermionic
vector  field will induce a change in the dilaton $\phi$ by
\eqn\dilaton{\phi' = \phi + \half \log B_{11}.}
since the integration of the vector field has exactly the same formal form as the one
we had for the bosonic T-duality, except that in this case we are integrating
over an anticommuting
variable. Thus,
 the change in $\phi$ under fermionic T-duality has the
opposite sign from the change in $\phi$ under bosonic T-duality.
Another difference with bosonic T-duality is that fermionic
T-duality does not change the relative sign of $\pb\t^1/\pb\widetilde \theta^1$ versus
$\p \t^1/\p\widetilde \theta^1$, and does not change the relative sign of
$L'_{1M}/L_{1M}$ versus $L'_{M 1}/L_{M1}$.
We can find the explicit on-shell relation between the original and the T-dualized variables
by computing the equations of motion for $(A,\bar A)$ and using the equation of motion for
the field $\widetilde \theta^1$ which implies that the vector field is given by the gradient
of $\theta^1$. We find
\eqn\differm{\pb \widetilde \theta^1 = B_{11} \bar \partial \theta^1  -
(-1)^{s(M)}L_{1M}\pb Y^M,
\quad
\p \widetilde \theta^1 = B_{11} \partial  \t^1 + L_{M1}\p Y^M,}
where $s(M) =0$ if $M$ is bosonic and $S(M)=1$ if $M$ is fermionic.
On the other hand the equations that relate a boson to the T-dual boson, coming from \actionbg ,
 are
\eqn\differbm{\pb \widetilde x^1 = -( g_{11}\pb x^1 + l_{1\mh}\pb x^\mh),
\quad
\p  \widetilde x^1 = g_{11}\p x^1 + l_{\mh 1}\p x^\mh.
}
In other words, we have $d \widetilde x^1 = g_{11} * d x^1 + \cdots $ for the boson while
we have $d \widetilde \theta^1 = B d \theta^1 + \cdots $ for the fermion. Notice the absence of
the $*$ for the fermionic case.

  Note that the fermionic variables are morally non-compact. Our arguments here have
  ignored the fact that the vector field can have non-trivial holonomies on the Riemann surface.
  Thus our derivation is only justified in the case of the disk but not on
  higher genus Riemann surfaces. Even on   the disk, we will need to
  treat the zero modes of the original and the T-dual fermion in a special way. We will
  integrate over the zero modes of the initial fermion before doing the T-duality and
  we will not integrate over the fermion zero modes of the T-dual fermions.
(This is similar to the treatment of non-compact bosonic zero modes on the  disk.)

If one wanted to define fermionic T-duality on a higher genus Riemann
surface, one would need to introduce fermionic variables which are allowed
to be non-periodic when its worldsheet location $z$ is
taken around a non-trivial cycle on the surface.
Note that the usual Green-Schwarz $\t$ variables are defined to be periodic and satisfy
\eqn\periodic{\t(z +C_i) = \t(z)}
where $C_i$ is any non-trivial cycle on the worldsheet.
If one wants to require that the fermionic vector field $(A,\bar A)$
has trivial holonomies
so that it can be expressed as the gradient of $\t$,
one would need to use a Lagrange multiplier term
$\int d^2 z~\tt (\p \bar A - \pb A)$
where $\tt(z)$ is a non-periodic variable satisfying
\eqn\nonper{\tt(z + C_i) = \tt(z) + \rho_i,}
and $\rho_i$ are Grassmann constants which need to be integrated over.

So if the original fermionic variable is periodic, the dual fermionic
variable is non-periodic and contains an extra zero mode for every non-trivial
cycle on the worldsheet. Similarly, if the original fermionic variable
is non-periodic, the dual fermionic variable will be periodic and
the holonomies of the vector field around the non-trivial cycles
will correspond to the $\rho_i$ constants in \nonper.
This T-dual relation between periodic and non-periodic fermionic variables
is analogous to the T-dual relation between non-compact bosonic variables
and bosonic variables compactified on a circle of zero radius.

\subsec{T-duality in pure spinor formalism}

Although one normally does not expect two-derivative terms for fermions such
as $\int d^2 z B_{11}\p\t^1\pb\t^1$, these terms arise in Green-Schwarz
and pure spinor
sigma models for Type II superstrings in Ramond-Ramond backgrounds.
To find how the T-duality transformations of \primefields\ act
on the Type II supergravity background fields, one needs to know the relation
of $L_{MN}(Y)$ with the onshell supergravity fields. In the Green-Schwarz
formalism, this relation depends on the choice of superspace torsion
constraints and can be quite complicated.
As recently discussed in \policastro, the most convenient method
for determining this relation is to use the pure spinor formalism
where BRST invariance determines the choice of torsion constraints
and allows a straightforward identification of the background fields.

In the pure spinor version of the Type II sigma model, the worldsheet
action is
\eqn\pureaction{
{1\over{2\pi\a'}}
\int d^2 z [L_{MN}(Z) \p Z^M \pb Z^N + P^{\a\bh}(Z) d_\a \hat d_\bh
+ E_M^\a(Z) d_\a \pb Z^M + E^\ah_M(Z) \p Z^M \hat d_\ah  }
$$+ \Omega_{M\a}^\b(Z)\l^\a w_\b \pb Z^M + \hat\Omega_{M\ah}^\bh (Z)\p Z^M \lh^\ah \hat w_\bh
+ C_{\a}^{\b\gh}(Z)\l^\a w_\b \hat d_\gh + \hat C_{\ah}^{\bh\g} (Z) d_\g \lh^\ah \hat w_\bh$$
$$+S_{\a\gh}^{\b\dh}(Z)\l^\a w_\b \lh^\gh \hat w_\dh +
w_\a \pb \l^\a + \hat w_\ah \p \l^\ah] + {1\over{4\pi}}\int d^2 z \Phi(Z) {\cal R}
$$
where $Z^M$
are coordinates for N=2 d=10 superspace, $d_\a$ and $\hat d_\ah$ are
independent fermionic variables, $(\l^\a,w_\a)$ and
$(\lh^\ah,\hat w_\ah)$ are the left and right-moving pure
spinor ghosts, and ${\cal R}$ is the worldsheet curvature.
BRST invariance implies relations between the various superfields
appearing in \pureaction\ where the BRST operators
are $Q=\int dz ~\l^\a d_\a$ and $\hat Q = \int d\bar z
~\lh^\ah \hat d_\ah$. By comparing with the vertex operators
for massless fields, one learns that the $\t=\hat\t=0$ component
of $P^{\a\bh}$ is $P^{\a \bh}|_{\t=\hat \t =0}  = -{ i \over 4}  e^{\phi} F^{\a\bh}$
 where $F^{\a\bh}$
is the Ramond-Ramond field
strength in bispinor notation\foot{ The relation to the usual notation for the RR field strengths of
type IIB string theory is $
F^{\alpha \hat \beta} = (\gamma^m)^{\alpha \hat \beta} F_m + { 1 \over 3 !}
(\gamma^{ m_1 m_2 m_3})^{\alpha \hat \beta} F_{m_1 m_2 m_3} + { 1 \over 2} { 1 \over 5!}
 (\gamma^{ m_1\cdots m_5})^{\alpha \hat \beta} F_{m_1 \cdots  m_5} $. The factor of $e^\phi$
in $P= -{ i \over  4 } e^\phi F$ is present since $P$ has the kinetic term $\int d^{10} x e^{-2\phi} P^2.$ }, the $\t=\hat\t=0$ components
of $E_m^\a$ and $E_m^\ah$ are the N=2 d=10 gravitinos,
the $\t=\hat\t=0$ components of
$\Omega_{m\a}^{\b} (\g^{ab})^\a_\b \pm \hat \Omega_{m\ah}^{\bh}
(\g^{ab})^\ah_\bh$ are the spin connection and NS-NS three-form,
the $\t=\hat\t=0$ components
of $C_{\a}^{\b\gh} (\g^{ab})^\a_\b$ and $\hat C_{\ah}^{\bh\g} (\g^{ab})^\ah_\bh$
are the N=2 gravitino field-strengths, and
the $\t=\hat\t=0$ component of
$S_{\a\gh}^{\b\dh} (\g^{ab})^\a_\b (\g^{cd})^\gh_\dh$ is the Riemann tensor
and the derivative of the NS-NS three-form.

If \pureaction\ is invariant under the fermionic shift in  \shift,
one can easily apply the Buscher procedure of the previous subsection
to the action of \pureaction. One finds that \pureaction\ is T-dualized
to
\eqn\tpure{\eqalign{
& {1\over{2\pi\a'}}
\int d^2 z [B'_{11}(Y) \p \tto \pb \tto +
L'_{1M}(Y) \p\tto\pb Y^M + L'_{M1}(Y)\p Y^M\pb \tto + L'_{MN}\p Y^M \pb Y^N +
\cr
& + P'^{\a\bh}(Y) d_\a \hat d_\bh
+ E'^\a_1(Y) d_\a \pb \tto  +  E'^\a_M(Y) d_\a \pb Y^M
+E'^\ah_1(Y)\p\tto \hat d_\ah  + E'^\ah_M(Y) \p Y^M \hat d_\ah + ...]
\cr
&+{1\over{4\pi}}\int d^2 z \Phi'(Y) {\cal R}
}}
where
$Y^M$ ranges over all bosonic and fermionic variables except for
$\t^1$, the superfields
$[B'_{11},L'_{1M},L'_{M1},L'_{MN}, \Phi']$ are defined as in
\primefields, and
\eqn\newprime{P'^{\a\bh} = P^{\a\bh} - (B_{11})^{-1} E_1^\a E_1^\bh,\quad
E'^\a_1 = (B_{11})^{-1} E^\a_1, \quad
E'^\ah_1 = (B_{11})^{-1} E^\ah_1, }
$$E'^\a_M = E^\a_M - (B_{11})^{-1} L_{1M} E_1^\a,\quad
E'^\ah_M = E^\ah_M - (B_{11})^{-1}  E_1^\ah L_{M1},$$
$$\Omega'^\b_{1\a} = (B_{11})^{-1} \Omega_{1\a}^\b, \quad
\hat{\Omega'}^\bh_{1\ah} = (B_{11})^{-1} \hat\Omega_{1\ah}^\bh, $$
$${\Omega'}_{M\a}^\b = \Omega_{M\a}^\b - (B_{11})^{-1} L_{1M}
\Omega_{1\a}^\b,\quad
\hat{\Omega'}_{M\ah}^\bh = \hat\Omega_{M\ah}^\bh - (B_{11})^{-1} \hat
\Omega_{1\ah}^\bh L_{M1},$$
$$C'^{\b\gh}_\a = C^{\b\gh}_\a - (B_{11})^{-1} E_1^\gh\Omega_{1\a}^\b,
\quad
\hat {C'}^{\bh\g}_\ah = \hat C^{\bh\g}_\ah - (B_{11})^{-1} \hat\Omega_{1\ah}^\bh E_1^\g,$$
$$S'^{\b\dh}_{\a\gh} = S^{\b\dh}_{\a\gh} - \hat\Omega_{1\gh}^\dh
\Omega_{1\a}^\b.$$

Note that the worldsheet variables
in the BRST operators $Q=\int dz \l^\a d_\a$ and $\hat Q = \int d\bar z
\lh^\ah \hat d_\ah$ are not affected by fermionic
T-duality, so BRST invariance is manifestly
preserved.
Although the fermionic T-duality transformations of \newprime\ are similar to the bosonic
T-duality transformations discussed in \policastro,
there are some crucial differences. For example,
$E'^\a_1$ has the same relative sign as $E'^\ah_1$ in\newprime.
But in bosonic T-duality
if one dualizes the $x^p$ coordinate as in \policastro,
\eqn\bostd{E'^\a_p = (G_{pp})^{-1} E^\a_p, \quad
E'^\ah_p = - (G_{pp})^{-1} E^\ah_p.}
As will now be explained, this difference
implies that unlike bosonic T-duality,
fermionic T-duality does not exchange the Type IIA and Type IIB superstrings
and does not modify the dimension of the D-brane.

As discussed in \howeb,
the pure spinor Type II sigma model and BRST operators are invariant under three independent
local Lorentz transformations which transform
\eqn\dualityf{\d E^a_M = L^a_b E^b_M, \quad \d E^\a_M = M^{ab} (\g_{ab})^\a_\b E^\b_M,
\quad \d E^\ah_M = \hat M^{ab} (\g_{ab})^\ah_\bh E^\bh_M,}
$$\d d_\a = M^{ab} (\g_{ab})_\a^\b d_\b,
\quad \d \hat d_\ah = \hat M^{ab} (\g_{ab})_\ah^\bh \hat d_\bh, \quad ...$$
where $M^{ab}$ and $\hat M^{ab}$ are independent of $L^{ab}$ and $...$ denotes similar
transformations on all background fields and worldsheet fields with tangent-space spinor
indices.
Furthermore, it was shown in \howeb\ that BRST invariance of the sigma model implies
the superspace torsion constraints
\eqn\torp{T^a_{\a\b} =i~ f^a_b~ \g^b_{\a\b},\quad
T^a_{\ah\bh} = i~\hat f^a_b~ \g^b_{\ah\bh},}
where $f^a_b$ and $\hat f^a_b$ are $O(9,1)$ matrices.

To compare with the usual description of Type II supergravity which has the
torsion constraints
\eqn\standt{T^a_{\a\b}=i~\g^a_{\a\b}, \quad
T^a_{\ah\bh}=i~\g^a_{\ah\bh},}
 one can use the local Lorentz symmetries
of $M^{ab}$ and $\hat M^{ab}$ to gauge-fix $f^a_b$ and $\hat f^a_b$.
After gauge-fixing,
only the combined  Lorentz symmetry of all three types of indices together
is preserved,   which
is the usual local Lorentz symmetry of supergravity.
If $f^a_b$ and $\hat f^a_b$ are $SO(9,1)$ matrices
with determinant $+1$, one can gauge $f^a_b = \hat f^a_b = \delta^a_b$
and recover \standt. But to recover \standt\ when $f^a_b$ (or
$\hat f^a_b$) has determinant $-1$, one needs to flip the
chirality of the unhatted (or hatted) spinor.

After performing bosonic T-duality (say in a flat background)
with respect to the coordinates $(x^1,...,x^p)$,
the relative minus sign in the transformation of $E_M^\a$ versus $E_M^\ah$
in \bostd\ implies
that the components $(f^1_1, ..., f^p_p)$ of $f^a_b$ have opposite sign with
respect to the components
$(\hat f^1_1, ..., \hat f^p_p)$ of $\hat f^a_b$.
So to return to the standard torsion constraints of \standt,
one needs to perform local Lorentz transformations using $M^{ab}$ and $\hat M^{ab}$
which cancel this change in relative sign in $f$ versus $\hat f$.
These local Lorentz transformations
modify in the expected manner the
D-brane boundary conditions
which relate hatted and unhatted spinors. Furthermore,
if $p$ is odd, the determinants of $f$ and $\hat f$
will have opposite sign. So
to recover the torsion constraints of \standt,
one will have to flip the chirality of either the hatted or
unhatted spinors,
which switches the Type IIA and Type IIB superstring.

On the other hand, since
in fermionic T-duality there are no relative minus signs in the transformation
of $E_M^\a$ versus $E_M^\ah$, one does not need to perform local Lorentz
rotations to return to the constraints of \standt. So there is no switch of Type IIA
and Type IIB superstrings, and no modification of the dimension of
the D-brane.

\subsec{Transformations of component fields}

By considering the $\t=\th=0$ components of
the superfields in \newprime, one finds that
the fermionic T-duality transformations
leave invariant the NS-NS fields $g_{mn}$ and $b_{mn}$, and transform the
Ramond-Ramond bispinor field-strength $F^{\a\bh}$ and
dilaton $\phi$ as
\eqn\phit{-{ i \over 4 } e^{\phi'}F'^{\a\bh} = -{ i \over 4 }
e^{\phi} F^{\a\bh} - \e^\a \eh^\bh C^{-1},\quad
\phi' = \phi +\half \log C}
where $C$ is the $\t=\th=0$ component of $B_{11}$ and
$(\e^\a,\eh^\ah)$ are the $\t=\th=0$ components of $(E_1^\a,E_1^\ah)$.
Although it is not difficult to also work out the T-duality
transformations of the fermionic fields, we will assume here
that all fermionic background fields have been set to zero.

To determine the relation of $C$ and $(\e^\a,\e^\ah)$
with the supergravity fields, note that the torsion constraints
imply that the superspace
3-form field-strength
\eqn\threef{H_{ABC} = E_A^M E_B^N E_C^P H_{MNP} =
 E_A^M E_B^N E_C^P \p_{[M} B_{NP]} }
has constant spinor-spinor-vector components \Howewest
\eqn\consts{H_{\a\b c} =i (\g_c)_{\a\b},\quad H_{\ah\bh c} = -i(\g_c)_{\ah\bh},
\quad H_{\a\bh c}=0,}
where $A=(c,\a,\ah)$ denotes tangent-superspace indices, $M$ denotes
curved-superspace indices, and $E_A^M$ is the inverse super-vierbein.
(The relative minus sign in $H_{\a\b c}$ versus $H_{\ah\bh c}$ is because
$H\to - H$ under a worldsheet
parity transformation which switches $z\to\bar z$ and
$\a\to\ah$.)

Since the fermionic isometry implies that $\p_1 B_{1m}=0$ where
$\p_1 \equiv {\p\over{\p\t^1}}$, one finds that
\eqn\derivc{\eqalign{\p_m C = & \p_m B_{11}|_{\t=\th=0} = H_{11 m}|_{\t=\th=0}
= E_1^A E_1^B E_m^C H_{ABC}|_{\t=\th=0}
\cr= & i \e^\a \e^\b e_m^c (\g_c)_{\a\b} -i \eh^\ah \eh^\bh e_m^c (\g_c)_{\ah\bh} =
i\e \gamma_m \e - i\eh \gamma_m \eh
}}
where $e_m^c \equiv {E_m^c}|_{\t=\th=0}
$ is the usual vierbein, $E_1^\a|_{\t=\th=0} =
\e^\a$ and $E_1^\ah|_{\t=\th=0} = \eh^\ah$.

Under the fermionic isometry
of \shift,
\eqn\fermi{E^\a_M \d Z^M = E^\a_1 \rho , ~~~~~
\quad E^\ah_M \d Z^M = E^\ah_1\rho
}
where $\rho$ is a constant anticommuting parameter. Since the
$\t=\th=0$ components of $E_M^\a \d Z^M$ and
$E_M^\ah \d Z^M$ are the local supersymmetry parameters \wessbagger,
and since
$E^\a_M \d Z^M|_{\t=\th=0} = \e^\a \rho $
and $E^\ah_M \d Z^M|_{\t=\th=0} = \hat\e^\ah \rho $,
the isometry
of \shift\ implies that
the component background supergravity fields are invariant under the
   supersymmetry
transformation parameterized by the Killing
spinors $\e^\a = E_1^\a|_{\t=\th=0}$ and $\eh^\ah=
E_1^\ah|_{\t=\th=0}$. Note that we are talking about one supersymmetry given
by these two spinors, and not two independent supersymmetries.
So \derivc\ implies that the derivative of $C$ is related to the
Killing spinors $\e^\a$ and $\eh^\ah$.
Note that the constant part of $C$ is unconstrained, as
can be seen from the fact that $B_{11}\p\t^1 \pb\t^1$ changes
by a total derivative under a constant shift of $B_{11}$.

Since the fermionic isometry is assumed
to be abelian (i.e. $Q^2 =0$), one learns from the supersymmetry algebra
\eqn\susya{(\e^\a Q_\a + \eh^\ah Q_\ah)^2 =
(\e\g^m \e + \eh\g^m\eh) P_m}
that
\eqn\susyreal{\e\g^m\e + \eh\g^m\eh=0}
where $(Q_\a,Q_\ah)$ are the supersymmetry generators and $P_m$
is the translation generator.
So \derivc\ implies that $\p_m C = 2i\e\g_m\e = -2i\eh\g_m\eh$.
Note that
if $\e^\a$ and $\eh^\ah$ were Majorana spinors, \susyreal\ would imply
that $\e^\a=\eh^\ah=0$ since $(\gamma^0)_{\alpha \beta}$
is equal to the identity matrix in this basis.
So the only non-trivial solutions to \susyreal\
involve complex Killing spinors $\e^\a$ and $\eh^\ah$.
In general, the T-duality transformation of \phit\ will therefore
not map real background fields into real background fields.

\subsec{Supersymmetry of T-dualized background}

As was shown in the previous subsection, the fermionic T-duality
transformation of \primefields\ and \dilaton\ leaves invariant the component
NS-NS fields $g_{mn}(x)$ and $b_{mn}(x)$, and transforms the
Ramond-Ramond bispinor field-strength $F^{\a\bh}(x)$ and
dilaton $\phi(x)$ as
\eqn\phit{-{ i \over 4 } e^{\phi'}F'^{\a\bh} =
-{ i \over 4 }  e^{\phi} F^{\a\bh} - \e^\a \eh^\bh C^{-1},\quad
\phi' = \phi +\half \log C}
where $C(x)$ is the $\t=\th=0$ component of $B_{11}$ which
satisfies
\eqn\ccdeftwo{\p_m C = 2i\e\g_m\e = -2i\eh\g_m\eh,}
and
$(\e^\a(x),\eh^\ah(x))$ are the Killing spinors associated
to the fermionic shift isometry of \shift. In other words,
if one performs a local Type II supersymmetry transformation with
Killing spinors $(\e^\a(x),\eh^\ah(x))$, the original background is
assumed to be invariant.

A useful check of the transformations of \phit\ is that
they should map a supersymmetric Type II background into
a supersymmetric Type II background. If the original supersymmetry
corresponding to a constant shift of $\t^1$ is described by
Killing spinors $(\e,\eh)$, the T-dualized supersymmetry corresponding
to a constant shift of $\tto$ will be described by Killing spinors
$\e' = C^{-1}\e$ and $\eh' = C^{-1} \eh$.
One can also consider backgrounds with $n$ abelian supersymmetries
corresponding to constant shifts of $\t^J$ for $J=1$ to $n$.
In this case, the $n$ Killing spinors $(\e_J^\a, \eh_J^\ah)$
should satisfy the identities
\eqn\nabel{\e_J^\a \g^m_{\a\b} \e_K^\b + \eh_J^\ah \g^m_{\ah\bh}
\eh_K^\bh = \e_J  \g^m  \e_K  + \eh_J  \g^m
\eh_K =0 }
for $J,K=1$ to $n$ so that the $n$ supersymmetries anticommute
with each other.

After performing T-duality with respect to
$\t^J$ for $J=1$ to $n$, one finds that the Ramond-Ramond
field-strength $F^{\a\bh}(x)$ and
dilaton $\phi(x)$ transform as
\eqn\phitn{-{ i \over  4 } e^{\phi'}F'^{\a\bh} =
-{ i \over 4 } e^{\phi} F^{\a\bh} - \e_J^\a  (C^{-1})_{JK} \eh_K^\bh,\quad
\phi' = \phi +\half \sum_{J=1}^n (\log C)_{JJ}}
where $C_{JK}(x) = C_{KJ}(x)$ is the $\t=\th=0$ component of
$B_{JK}$ which
satifies
\eqn\ccdeftwon{\p_m C_{JK} = 2i\e_J\g_m\e_K = -2i\eh_J\g_m\eh_K.}
Furthermore, the new Killing spinors after performing T-duality
are
\eqn\newk{{\e'}_J^\a = (C^{-1})_{JK} \e_K^\a, \quad
\eh'{}^\ah_J = (C^{-1})_{JK} \e_K^\ah.}

Under N=2 d=10
supersymmetry transformations parameterized by
$(\rho_J \e_J^\a, \rho_J \eh_J^\a)$ where
$\rho_J$ are Grassmann constants,
the dilatino $\l_\a$ and gravitino $\psi^\a_m$
transform in string frame as \BergshoeffPV \foot{Our conventions differ from the
ones in \BergshoeffPV\ by  a factor of 4 for the RR fields. Namely, we have
$ P = -{ i \over 4} e^\phi F^{\alpha \hat \beta}_{ours} =
-{ i \over 16 }e^\phi F^{\alpha \hat \beta}_{\BergshoeffPV} $ where  $
F^{\alpha \hat \beta} = (\gamma^m)^{\alpha \hat \beta} F_m + { 1 \over 3 !}
(\gamma^{ m_1 m_2 m_3})^{\alpha \hat \beta} F_{m_1 m_2 m_3} + { 1 \over 2} { 1 \over 5!}
 (\gamma^{ m_1\cdots m_5})^{\alpha \hat \beta} F_{m_1 \cdots  m_5} $ in both cases.}
\eqn\susytr{\d_J \l_\a = \p_m\phi (\g^m\e_J)_\a +
2 i(\g_m P \g_m \eh_J)_\a + { 1 \over 12} H_{mnp} (\g^{mnp}\e_J)_\a,}
$$\d_J\psi_m^\a =
\n_m\e_J^\a + 2i (P \g_m \eh_J)^\a + { 1 \over 8 } H_{mnp} (\g^{np}\e_J)^\a,$$
where $P^{\a\bh} \equiv -  {i\over 4}
e^{\phi} F^{\a\bh}$
and $H_{mnp}$
is the Neveu-Schwarz three-form field-strength.
After T-dualizing all fields and Killing spinors on the
right-hand side of \susytr, one finds
\eqn\susytrtwo{\eqalign{ \d'_J \l'_\a =&
\p_m\phi' (\g^m\e'_J)_\a + 2 i(\g_m P' \g^m \eh'_J)_\a
                          + { 1 \over 8} H_{mnp} (\g^{mnp}\e'_J)_\a
\cr = & (C^{-1})_{JK} \d_K\l_\a +{1\over 2}
(C^{-1})_{KL} (\p_m C)_{LK} (\g^m \e_M)_\a (C^{-1})_{JM}
 \cr
& - 2i(\g^m\e_K)_\a (C^{-1})_{KL} (\eh_L \g^m \eh_M)(C^{-1})_{JM}
 \cr =&
(C^{-1})_{JK} \d_K\l_\a +
i(C^{-1})_{KL} (\e_L\g_m\e_K)  (\g^m \e_M)_\a (C^{-1})_{JM} +
\cr
+ & 2i (\g^m\e_K)_\a (C^{-1})_{KL} (\e_L \g^m \e_M)(C^{-1})_{JM}
 =
(C^{-1})_{JK} \d_K\l_\a }}
where we used the gamma-matrix identity
\eqn\gide{(\e_L\g_m \e_K)(\g^m\e_J)_\a +
(\e_K\g_m \e_J)(\g^m\e_L)_\a +
(\e_J\g_m \e_L)(\g^m\e_K)_\a =0}
 We also have
\eqn\susytrth{ \eqalign{
&\d'_J\psi'^\a_m =    \n_m{\e'}_J^\a + 2i (P' \g_m \eh'_J)^\a +
{ 1 \over 12 } H_{mnp} (\g^{np}\e'_J)^\a
\cr = &  (C^{-1})_{JK} \d_K\psi_m^\a -
 (C^{-1}(\p_m C )C^{-1})_{JK} \e_K^\a
-2i \e_K^\a (C^{-1})_{KL} (\eh_L\g_m \eh_M) (C^{-1})_{JM}\cr
=& (C^{-1})_{JK} \d_K\psi_m^\a + 2i
 (C^{-1})_{JM} (\eh_M\g_m \eh_L)(C^{-1})_{LK} \e_K^\a
-2i \e_K^\a (C^{-1})_{KL} (\eh_L\g_m \eh_M) (C^{-1})_{JM}
\cr
 = & (C^{-1})_{JK} \d_K\psi_m^\a
 }}

So if the background is supersymmetric before T-duality (i.e.
if $\d_J\l_\a = \d_J\psi_m^\a =0$), it is also
supersymmetric after T-duality (i.e. $\d'_J\l'_\a = \d'_J
{\psi'}_m^\a=0$).

\subsec{Null Ramond-Ramond field strength}

The simplest example of fermionic T-duality is in a flat background
where the supersymmetry parameters $\e^\a$ and $\eh^\bh$ are
constants. One usually does not include the term $\int d^2 z B_{11}\p\t^1
\pb\t^1$ in the flat worldsheet action, but if $B_{11}$ is constant,
this term is a total derivative and can be included without affecting
the equations of motion.

Since $\p_m B_{11}=0$, \ccdeftwo\
implies that the supersymmetry parameters must be
chosen to satisfy
\eqn\pureco{\e\g^m\e=\eh\g^m\eh=0,}
i.e. $\e^\a$ and $\eh^\ah$
are d=10 pure spinors. Since \pureco\ has no Majorana-Weyl solutions
in d=10 Minkowski space, one needs to consider complexified
supersymmetry parameters.

After performing the T-duality transformations of
\phit,
one finds that the dilaton shifts by a constant
and the Ramond-Ramond field strength picks up the constant value
\eqn\rrf{e^{\phi'}{F'}^{\a\bh} = 4i\e^\a \eh^\bh C^{-1}.}
Since the stress tensor $T^{mn}$ for a bispinor Ramond-Ramond field strength
is proportional to $\g^m_{\a\b}\g^n_{\gh\dh} F^{\a\gh} F^{\b\dh}$ and
since
$\e^\a$ and $\eh^\bh$ are pure spinors satisfying \pureco,
${F'}^{\a\bh}$ is a ``null'' bispinor which does not contribute
to the stress-tensor
and does not produce a back-reaction.

A closely related example which will
be discussed in the following subsection arises as follows.
One starts with a Calabi-Yau compactification
to four dimensions which preserves N=2 d=4 supersymmetry,
and one chooses $\e^\a$ and $\eh^\bh$ to be
the chiral N=2 d=4 supersymmetry
parameters. In this case, the resulting T-dualized background
of \rrf\ involves
the self-dual graviphoton field-strength of \refs{\oog,\seibergone,\seiberg}
which leads
to non-anti-commutative N=1 d=4 super-Yang-Mills on a D3 brane.
As predicted by T-duality, the closed superstring spectrum
in this self-dual graviphoton background is identical to
the spectrum without the self-dual graviphoton field-strength.
But this
example shows clearly that fermionic T-duality changes the theory at higher loops since, unlike in a flat background,
certain $F$ terms in the effective action
 for a constant graviphoton background have been computed \BCOV\ and are non-zero in general.

\subsec{Self-dual graviphoton background}

To explicitly derive the T-duality transformations for
the sigma model in a flat $d=4$ background with Calabi-Yau
compactification,
it is convenient to use the $d=4$ hybrid
formalism for describing the worldsheet action. In a flat
background, the worldsheet action is
\eqn\hybr{S = \int d^2 z [\p x^{a\ad} \pb x_{a\ad}
 + p_a \pb\t^a + \hat p_a\p\th^a + \bar p_\ad \pb\tb^\ad
+ \hat{\bar p}_\ad\p\hat{\tb}^\ad] + S_C}
where $a,\ad=1$ to 2 and $S_C$ is the action for the compactified sector
of the superstring. As discussed in \seibergone, one can choose
a chiral representation such that $q_a=\int dz p_a$ and $\hat q_a=
\int d\bar z \hat p_a$
are the chiral spacetime supersymmetry generators. In this chiral
representation, both the worldsheet action and the BRST operator
are invariant under the shift isometries
\eqn\isotwo{\t^a \to \t^a + \rho^a, \quad \th^a \to \th^a + \hat\rho^a}
where $\rho^a$ and $\hat\rho^a$ are constants and all other
worldsheet variables are unchanged.

After adding to \hybr\ the surface term
\eqn\surf{\int d^2 z   C_{ab}[ \p\t^a\pb\th^b -   \pb\t^a\p\th^b ]}
where $C_{ab}=C_{ba}$ is a constant symmetric bispinor,
one can T-dualize the shift isometries of \isotwo\ by introducing
the fermionic gauge fields $(A^a,\bar A^a)$ and $(\hat A^a,\hat{\bar A}^a)$
to obtain the action
\eqn\hybrtwo{S = \int d^2 z [\p x^{a\ad} \pb x_{a\ad}
 + p_a \bar A^a + \hat p_a \hat A^a +
C_{ab}(A^a\hat{{\bar A^b}} -\bar A^a\hat A^b)}
$$ + \tt_a (\p\bar A^a -\pb A^a) +
\hat\tt_a (\p\hat{\bar A^a}  -\pb \hat A^a) + \bar p_\ad \pb\tb^\ad
+ \hat{\bar p}_\ad\p\hat{\tb^\ad} ] + S_C. $$

Integrating out the worldsheet gauge fields produces a constant shift
of the dilaton and the worldsheet action becomes
\eqn\hybrthree{S = \int d^2 z [\p x^{a\ad} \pb x_{a\ad}
 + (C^{-1})^{ab}(p_a \pb \hat \tt_b + \hat p_a\p  \tt_b + p_a \hat p_b
- \p\tt_a\pb\hat\tt_b
 +\pb\tt_a\p\hat\tt_b)}
$$
 + \bar p_\ad \pb\tb^\ad
+ \hat{\bar p}_\ad\p\hat{\tb^\ad} ] + S_C. $$
After dropping the surface term
$\int d^2 z (C^{-1})^{ab}( \p\tt_a\pb\hat\tt_b
 +\pb\tt_a\p\hat\tt_b )$ and defining $\phi^a = (C^{-1})^{ab}\hat \tt_b$
and
$\hat\phi^a = (C^{-1})^{ab} \tt_b$, one obtains the action
\eqn\hybrfive{S = \int d^2 z [\p x^{a\ad} \pb x_{a\ad}
 + p_a \pb\phi^a + \hat p_a\p\hat\phi^a + (C^{-1})^{ab} p_a \hat p_b
 + \bar p_\ad \pb\tb^\ad
+ \hat{\bar p}_\ad\p\hat{\tb^\ad} ] + S_C,}
which is the worldsheet action of
\refs{\oog,\seibergone,\seiberg} in
a background with constant
self-dual field-strength $F^{ab}$ proportional to $e^{-\phi} (C^{-1})^{ab}$.

The difference between loop amplitudes in the constant
self-dual graviphoton background
and loop amplitudes in a flat background comes from the presence of the term
$(C^{-1})^{ab}\int d^2 z p_a\hat p_b$ in the self-dual graviphoton worldsheet
action.
Since $p_a$ is holomorphic and $\hat p_a$ is antiholomorphic, this term
can be written as $(C^{-1})^{ab}(\int dz p_a) (\int d\bar z \hat p_b)$ where
the contours of $\int dz$ and $\int d\bar z$ go around the non-trivial cycles
of the genus $g$ surface. So this term can absorb the higher-genus zero modes
associated with the fermionic one-forms $p_a$ and $\hat p_a$. Again, higher genus
amplitudes are sensitive to the presence of the constant graviphoton field strength
\BCOV . This shows that fermionic T-duality is not a full symmetry of the theory at higher
genus.

\newsec{Exact T-Duality of the $AdS_5\times S^5$ Background}

In this section, we show
that after performing bosonic T-duality with respect to
the $d=4$ coordinates
$(x^0,x^1,x^2,x^3)$ and performing fermionic T-duality with respect
to 8 of the 32 fermionic coordinates $\t^{\a j}$, the original
$AdS_5\times S^5$
background is mapped to another $AdS_5\times S^5$ background with
constant dilaton.
The transformation is an exact change of variables in the path integral, with
a unit jacobian. Thus this is an exact symmetry to all orders in the $\alpha'$ expansion and
it is also expected to be an exact symmetry non-perturbatively in $\alpha'$.
We will first show this by analyzing the transformations of
the $AdS_5\times S^5$
 background fields, and we will then show it again by explicitly
T-dualizing the Green-Schwarz and pure spinor versions of the $AdS_5\times S^5$
sigma model.

\subsec{T-duality transformations of the $AdS_5\times S^5$ background fields}

A non-trivial example of fermionic T-duality  arises in the $AdS_5\times S^5$
background which has 32 fermionic isometries. These
isometries can be identified
with the N=4 d=4 supersymmetry transformations $(q^{aj}, \bar q^\ad_j)$
and the N=4 d=4 superconformal transformations $(s^a_j, \bar s^{\ad j})$,
and one can choose
8 of the 32 fermionic symmetries
to anticommute with each other and to also
commute with the four $d=4$ translations. A convenient choice for the abelian
subset are the 8 chiral supersymmetry generators $q^{aj}$ which
will be associated with the Killing spinors
$(\e_{aj}^\a, \eh_{aj}^\ah)$. After
T-dualizing with respect to these 8 abelian fermionic isometries,
\phitn\ implies that
\eqn\phitwo{-{i\over 4} e^{\phi'}F'^{\a\bh} =
-{i\over 4} e^{\phi} F^{\a\bh} - \e_{aj}^\a \eh_{bk}^\bh (C^{-1})^{aj~bk},\quad
\phi' = \phi +\half \Tr \log C}
We can determine $C$ in two ways. We could use the explicit form
of the Killing spinors and
use  \ccdeftwon, or we could view
  $C_{aj~bk}$ as  the $\t=\th=0$ component of $B_{aj~bk}$. We will follow this second route.

In an $AdS_5\times S^5$ background, one can choose a gauge where the
only nonzero components of $B_{AB} \equiv E_A^M E_B^N B_{MN}$ are
the components
\eqn\Bchoice{B_{\a\bh} = B_{\bh\a} = -i(\gamma^{01234})_{\a\bh}.}
This gauge choice is not possible in a flat background, and it
simplifies the Green-Schwarz Wess-Zumino term in an $AdS_5 \times S^5$ background \bershadsky.
In this gauge, $C_{aj~bk}$ is the $\t=\th=0$ component of
$\e^\a_{aj} \eh^\bh_{bk} B_{\a\bh} +
\e^\ah_{aj} \eh^\b_{bk} B_{\ah\b}$, so one finds that
\eqn\cads{C_{aj~bk} = -2i \e_{aj}^\a (\g^{01234})_{\a\bh} \eh_{bk}^\bh.}


It is convenient to write the $AdS_5\times S^5$ metric as
\eqn\metrica{ds^2 = |y|^{-2} (dx^m dx_m + dy_r dy_r) }
where ${{y_r}\over{|y|}}$ for $r=1$ to 6
are the variables on $S^5$ and $|y|$ is the fifth variable on $AdS_5$.
It is also convenient to decompose the local spinor indices $\alpha$, $\hat \alpha $
into $SO(3,1) \times SO(5)$ as $\a = (a'j', \ad' j')$. Note that $j'=1, \cdots ,4 $  is an $SO(5)$
spinor index that can be raised and lowered using $(\s^6)_{j'k'}$
where $(\s^r)_{j'k'}$ are the $SO(6)$ Pauli matrices. In terms of this decomposition
we have that $(\gamma^{01234})^{a' j' \, b' k' } = i  \epsilon^{a' b' } (\sigma^6)^{j' k' } $.
In order to write the form of the Killing spinors we introduce the rotation matrix $M^{k'}_{j}(y) $
which is the ${{SU(4)}\over{SO(5)}}$ matrix
which rotates the point
$(0,0,0,0,0,1)$ on $S^5$ to the point $|y|^{-1}(y_1,y_2,y_3,y_4,y_5,y_6).$
The
Killing spinors $\e_{aj}{}^\a$ and $\eh_{aj}{}^\ah$  can be written as
\eqn\susyps{\e_{aj}{}^{b'k'} = |y|^{\half}\d_a^{b'} M_j^{k'}(y),\quad
\e_{aj}{}^{\bd k'} =0,\quad
\eh_{aj}{}^{b'k'} = i |y|^{\half}\d_a^{b'} M_j^{k'}(y),
\quad \eh_{aj}^{\bd' k'}=0.}

Using \cads\ and the identity
\eqn\Mident{M^{j'}_l (\g^{01234})_{a'j'~b'k'} M^{k'}_m = i
\e_{a'b'} (\s^r)_{lm} |y|^{-1} y_r,}
one finds that
$C_{aj~bk} =   2i\e_{ab} \s^r_{jk} y_r$ and
$(C^{-1})^{aj~bk} = -{i\over 2}\e^{ab} (\s^r)^{jk} {{y_r}\over{|y|^2}}$. This formula
for $C$ obeys equation \ccdeftwon . In fact, we could have simply derived the expression
for $C$ by solving \ccdeftwon .
To determine the transformation of $F^{\a\bh}$ in \phitwo, note
that
\eqn\whichc{
\e_{aj}^{a'j'} (C^{-1})^{aj~bk} \eh_{bk}^{b'k'} =\half \epsilon^{a' b'} ( \sigma^6)^{j' k'} =
-{i\over 2} (\g_{01234})^{a'j'~b'k'}.}
Note that we get the projection of the matrix $\gamma^{01234}$ to the part with definite
four dimensional chirality. Thus, we can write it in terms of a projection operator
 $\half [(\g_{0123} - i)\g_4]^{\a\bh}$. This
only has nonzero components
when $\a = a'j'$ and $\bh = b'k'$ so that
$\half [(\g_{0123} - i)\g_4]^{a'j'~b'k'} =   (\g_{01234})^{a'j'~b'k'}$,
and one finds that
\eqn\phithree{\eqalign{ e^{\phi'} F'^{\a\bh} =&
 e^{\phi} F^{\a\bh} - 4i
\e_{aj}^\a \eh_{bk}^\bh (C^{-1})^{aj~bk} \cr
= &
(\g_{01234})^{\a\bh} -  (\g_{01234} - i\g_4)^{\a\bh} =
 (i\g_4)^{\a\bh}
}}
where the $\g$-matrices appearing in \phithree\ have tangent-space vector
indices. The dual background therefore
has an imaginary RR scalar field which varies only along
the radial $AdS$ direction.  Also, $Tr(\log C)= 8 \log |y|$ implies that
\eqn\dilf{\phi' = \phi +4 \log |y|.}

If one now T-dualizes with respect to the four translation
symmetries of $x^m$, it is easy to verify that the one-form Ramond-Ramond
field strength proportional to
$(i\g_4)^{\a\bh}$ transforms back into the five-form
Ramond-Ramond field strength proportional to $(\g_{01234})^{\a\bh}$, and that
the dilaton shifts back to $\phi = \phi' - 4 \log |y|$.
Note that the factor of $i$ in front of $\g^4$ disappears again in
Minkowski space when we
T-dualize along  the time direction $x^0$ \HullVG .  So the
$AdS_5\times S^5$ background fields are invariant under the combined
bosonic and fermionic T-duality transformations.

It is interesting to note that there is another combination of bosonic
and fermionic T-duality transformations which also leaves the $AdS_5\times S^5$
background invariant. If one breaks $SU(4)$ R-symmetry to
$U(1)\times SU(2)\times SU(2)$ by choosing a $U(1)$ direction in the $SU(4)$,
the $SU(4)$ index $j=1$ to 4 splits into an index $r=1$ to 2 which carries
$+1$ charge with respect to the chosen $U(1)$ direction,
and an index $r'=3$ to 4 which carries $-1$ $U(1)$
charge. Under this subgroup of $SU(4)$, the 32 supersymmetries split into
$(q^a_{r},q^a_{ r'}, \bar q_\ad^r,\bar q_\ad^{r'})$ and
$(s_a^{r},s_a^{ r'}, \bar s^\ad_r,\bar s^\ad_{r'})$, and one can choose
the
8 abelian supersymmetries to be $q^a_{r'}$ and $\bar q_\ad^{r}$ which
all carry $+1$ $U(1)$ charge. Furthermore, under the breakup of
$SU(4)$ into $U(1)\times SU(2)\times SU(2)$, the $SU(4)$ generators
$R_j^k$ split into $(R_r^s,R_{r'}^{s'},R_r^{s'}, R_{r'}^s)$
where the four generators $R_{r'}^{s}$ all carry $+2$ $U(1)$
charge.
Together with
the four translations of $(x^0,x^1,x^2,x^3)$, the 8 supersymmetries
$(q^a_{r'}, \bar q_\ad^{r})$ and
4 $SU(4)$ transformations $R_{r'}^{s}$ form an abelian subgroup of $PSU(2,2|4)$
isometries with 8 bosonic and 8 fermionic generators.
After performing T-duality with respect
to these 8 bosonic and 8 fermionic isometries, one finds using
a similar analysis as above that the $AdS_5\times S^5$
background is invariant.

Note that the translation generators $R_{r'}^s$
that we chose in the five-sphere are not hermitian, so
this choice will involve a complexification of the coordinates. An alternative way to see
this is to do an analytic continuation of the $S^5$ coordinates into $dS^5$ (five dimensional
de-Sitter space) and then write
the metric of $dS^5$ as
$ ds^2 = { - d w^2 + du_i du_i \over w^2 }$. With this choice, the four translation symmetries
shift the four $u$ coordinates.

This alternative choice of abelian isometries is related to harmonic
${\cal N}=4$ $d=4$ superfields in the same way that the previous choice of
abelian isometries using $q_{aj}$ is related to
chiral ${\cal N}=4$ $d=4$ superfields.
As discussed in \howeproj\ and
\siegelproj,
harmonic
${\cal N}=4$ $d=4$ superfields are naturally constructed using
the supercoset ${{PSU(2,2|4)}\over{PS(U(2|2)\times U(2|2))}}$
where the denominator $PS(U(2|2)\times U(2|2))$ consists of the
generators
\eqn\denomg{[M_a^b, M_\ad^\bd, D, R_r^s, R_{r'}^{s'}, q^a_r,
\bar q_\ad^{r'}, s_a^r,\bar s_{r'}^\ad].}
The
16 bosonic and 16 fermionic generators in the supercoset
${{PSU(2,2|4)}\over{PS(U(2|2)\times U(2|2))}}$
split into ``upper-triangular'' generators $[P^a_\ad,R_{r'}^s,
q_{r'}^a, \bar q_\ad^r]$ and ``lower-triangular''
generators
$[K_a^\ad,R^{r'}_s,
s^{r'}_a, \bar s^\ad_r]$,
and the ``upper-triangular'' generators are
precisely the $8+8$ abelian isometries which are T-dualized in this approach.
 This is closely related to the decomposition of $PSU(2,2|4)$ that one performs when 
 we consider a 1/2 BPS string state with large charge (corresponding to an operator $Tr[Z^J]$). 
 The upper vs. lower triangular generators act as creation vs annihilation operators for
 impurities along the string. 

\subsec{ Invariance of $AdS_5\times S^5$ Green-Schwarz sigma model}

This invariance under the combined fermionic and  bosonic
T-dualities can also be verified by explicitly performing
the T-duality transformations on the $AdS_5\times S^5$ sigma model.
To show this invariance,
we will first consider the Green-Schwarz version of the sigma model
and will then consider the pure spinor version.

In an $AdS_5\times S^5$ background, the Green-Schwarz sigma model
$S= \int d^2 z [ (G_{MN}(Z) + B_{MN}(Z))\p Z^M \bar\p Z^N]$ takes the
form
\eqn\gsaction{S = {R^2 \over 4 \pi \alpha' } \int d^2 z [ \eta_{cd} J^c \bar J^d
-i (\g^{01234})_{\a\bh} (J^\a \bar J^\bh - \bar J^\a J^\bh)]}
where $R$ is the $AdS$ radius,
$J^C = (g^{-1}\p g)^C$ and $\bar J^C = (g^{-1}\bar\p g)^C$ are
left-invariant Metsaev-Tseytlin \MetsaevIT\
currents constructed from the supercoset
$g(Z)\in {{PSU(2,2|4)}\over{SO(4,1)\times SO(5)}}$, and $C=(c,\a,\ah)$
labels the $PSU(2,2|4)$ Lie-algebra generators which are not in
$SO(4,1)\times SO(5)$. More precisely, $c=0$ to 4 labels the
five $AdS_5$ generators of ${{SO(4,2)}\over{SO(4,1)}}$, $c=5$ to 9 labels the
five $S^5$ generators of ${{SO(6)}\over {SO(5)}}$, $\a=1$ to 16 labels the
supersymmetries originating from the ``left-moving''
half of the ${\cal N}=2$ d=10 supersymmetry,
and $\ah=1$ to 16 labels the supersymmetries originating from the
``right-moving'' half.

Splitting the $SO(9,1)$ indices into $SO(3,1)\times SO(5)$ indices, this
action can be expressed as
\eqn\gsactionnew{\eqalign{ S = & {R^2 \over 4 \pi \alpha' } \int d^2 z [ (J_{P_m} +J_{K_m})(\bar J_{P_m} +
\bar J_{K_m}) + J_D \bar J_D +
\cr
 &~~~~~~~~
+  J_{R_t} \bar J_{R_t} +
J_{q^a_j} \bar J_{q^a_j}
+J_{\bar q_\ad^j} \bar J_{\bar q_\ad^j} +
J_{s_a^j} \bar J_{s_a^j} +
J_{\bar s^\ad_j} \bar J_{\bar s^\ad_j} ]}}
where $P_m$ and $K_m$ for $m=0$ to 3 label the
translations and conformal boosts, $D$ labels the dilatations,
$R_t$ for $t=1$ to 5 label the $SO(6)/SO(5)$ generators, and
$(q^a_j,\bar q_\ad^j, s_a^j,\bar s^a_j)$ label the fermionic
supersymmetry and superconformal generators.
Note that when written in terms of $SO(3,1)\times SO(5)$ spinor
indices, the $(\gamma^{01234})_{\a\bh}$ matrix
in \gsaction\ decomposes as $(\gamma^{01234})_{aj~bk}=i
\e_{ab}(\s^6)_{jk}$ and
$(\gamma^{01234})_{\ad j~\bd k}= i
\e_{\ad \bd}(\s^6)_{jk}$.
So
the $a$ and $\ad$ indices in \gsactionnew\ are contracted
with $\e_{ab}$ and $\e_{\ad\bd}$, while the $j$ indices are contracted
with $(\s^6)_{jk}$.

To compute the transformation of \gsactionnew\ under T-duality, it
is convenient to use the parameterization of the supercoset $g(Z)$
in which
\eqn\gsuperp{g(x^m, y^t,\t^{aj},\tb^\ad_j, \xb_\ad^j) =
\exp (x^m P_m + \t^{aj} q_{aj}) \exp (\tb^\ad_j \bar q_\ad^j +
\xb_\ad^j\bar s^\ad_j)~ |y|^D~
\exp(\sum_{t=1}^5 {{y^t}\over{|y|}} R_t)}
where
$|y|= \sqrt{\sum_{t=1}^6 y_t y_t}$
and ${{y_t}\over{|y|}}$ for $t=1$ to 6 are the $S^5$ coordinates.
In this parameterization of $g$, $\k$-symmetry has been used to gauge-fix
to zero the eight fermionic parameters associated with the ${\cal N}=4$ d=4
chiral superconformal generators $s^{aj}$. But there are still eight
remaining $\k$-symmetries which have not been gauge-fixed.

If one writes $g= \exp (x^m P_m + \t^{aj} q_{aj}) e^B$ where
\eqn\Bdef{e^B =
\exp (\tb^\ad_j \bar q_\ad^j +
\xb_\ad^j\bar s^\ad_j)~ |y|^D~
\exp(\sum_{t=1}^5 {{y^t}\over{|y|}} R_t),}
then the
left-invariant currents $g^{-1}\p g$ take the form
\eqn\bosl{J_{P_m} = [e^{-B}(\p x^n P_n +\p\t^{aj} q_{aj}) e^B]_{P_m},
\quad
J_{q^a_j} = [e^{-B} (\p x^m P_m +\p\t^{bk} q_{bk}) e^B]_{q^a_j}
}
$$
J_D = [e^{-B}\p e^B]_D,\quad J_{R_t} = [e^{-B}\p e^B]_{R_t}
,\quad J_{\bar q_\ad^j} =
[e^{-B} \p e^B]_{\bar q_\ad^j},\quad
J_{\bar s^\ad_j} =
[e^{-B} \p e^B]_{\bar s^\ad_j},$$
$$J_{K_m}=0, \quad J_{s_a^j} = 0,$$
where $[~~]_I$ denotes the component of $[~~]$ which is proportional
to the Lie-algebra generator $I$. To understand the structure of \bosl,
it is useful to note that the generators $(\bar q_\ad^j, \bar s^\ad_j,
D, R_j^k, M_{\ad\bd})$ form an $SU(2|4)$ supergroup where $M_{\ad\bd}$
are the anti-self-dual Lorentz generators. Under this $SU(2|4)$ supergroup,
the generators $(P_{a\ad},q_{aj})$ transform as a fundamental
representation and the generators
$(K^{a\ad},s^{aj})$ transform as an anti-fundamental representation.

One can now T-dualize with respect to $x^m$ and $\t^{aj}$ by introducing
the bosonic gauge fields $(A^m,\bar A^m)$ and the fermionic gauge fields
$(A^{aj}, \bar A^{aj})$, and adding the Lagrange multiplier term
\eqn\multip{{R^2 \over 4 \pi \alpha' } \int d^2 z [ \xt_m (\bar\p A^m -\p\bar A^m) + \tt_{a j}
(\bar\p
A^{aj} - \p \bar A^{aj})]}
to the action of \gsactionnew. The action then takes the form
\eqn\gsdualnew{\eqalign{S =& {R^2 \over 4 \pi \alpha' } \int d^2 z [ A'^m \bar A'^m + {A'}^{aj} {A'}^{aj} +
\xt_m (\bar\p A^m -\p\bar A^m) + \tt_{a j} (\bar\p
A^{aj} - \p \bar A^{aj}) + \cr &  ~~~~~~~~~~~ +  J_D \bar J_D + J_{R_t} \bar J_{R_t} +
J_{\bar q_\ad^j} \bar J_{\bar q_\ad^j} +
J_{\bar s^\ad_j} \bar J_{\bar s^\ad_j} ]
}}
where
\eqn\Aprime{A'^m = [e^{-B}(A^n P_n + A^{aj} q_{aj}) e^B]_{P_m}, \quad
{A'}^{aj} = [e^{-B}(A^m P_m + A^{bk} q_{bk}) e^B]_{q_{aj}}.}

Writing $A^m = [e^B ({A'}^n P_n + {A'}^{aj} q_{aj}) e^{-B}]_{P_m}$ and
$A^{aj} = [e^B ({A'}^n P_n + {A'}^{bk} q_{bk}) e^{-B}]_{q_{aj}}$ and integrating
out ${A'}^m$ and ${A'}^{aj}$, one finds that the T-dualized action is
\eqn\gsdualtwo{S = {R^2 \over 4 \pi \alpha' } \int d^2 z [ J'_{P_m} \bar J'_{P_m} +
{J'}_{q_{aj}} {\bar J'}_{q_{aj}}
+  J_D \bar J_D + J_{R_t} \bar J_{R_t} +
J_{\bar q_\ad^j} \bar J_{\bar q_\ad^j} +
J_{\bar s^\ad_j} \bar J_{\bar s^\ad_j} ]}
where $J'_{P_m} = [e^B (\p \xt_n P_m) e^{-B}]_{P_n} +
[e^B (\p \tt_{aj} P_m) e^{-B}]_{q_{aj}}$ and
${J'}_{q_{aj}} = [e^B (\p \xt_n q_{aj}) e^{-B}]_{P_n} +
[e^B (\p \tt_{bk} q_{aj}) e^{-B}]_{q_{bk}}$.

The integration over $A'$ and $\bar A'$ gives a measure factor
proportional to the superdeterminant of $|{{\p A'}\over{\p A}}|$.
Since $B$ is an element of $SU(2|4)$,
the super-Jacobian in the transformation of
\Aprime\ is equal to one. For example, if one
restricts to the dilatation
transformation parameterized by $|y|$, ${A'}^m = |y| A^m$ and
${A'}^{aj} = |y|^{\half} A^{aj}$. Since there are four $A^m$'s and
eight $A^{aj}$'s, the super-Jacobian cancels. So the measure factor is
equal to one which implies that the dilaton does not transform under
the combined bosonic and fermionic T-duality.

To relate \gsdualtwo\ to the original action of \gsactionnew, note that
\eqn\relj{J'_{P_m}
= Tr [e^B (\p\xt_n P_m) e^{-B} K^n + e^B(\p\tt_{aj} P_m) e^{-B}
s^{aj}] = [e^{-B} (\p\xt_n K^n + \p\tt_{aj} s^{aj}) e^B]_{K^m} }
where $Tr$ denotes the trace over $PSU(2,2|4)$ indices defined such that
$Tr(P_m K^n) = \delta_m^n$ and $Tr(q_{aj} s^{bk}) = \delta^a_b \delta_k^j$.
Similarly,
\eqn\relk{{J'}_{q_{aj}} = Tr [e^B (\p\xt_n q_{aj}) e^{-B} K^n + e^B(\p\tt_{bk} q_{aj})
e^{-B} s^{bk}] = [e^{-B} (\p\xt_n K^n + \p\tt_{bk} s^{bk}) e^B]_{s^{aj}}. }

Suppose one parameterizes
\eqn\paramnew{g(\xt,\tt,y,\tb,\xb)
= \exp (\xt_m K^m + \tt_{aj} s^{aj}) e^B}
where $e^B$ is defined as in
\Bdef\ and
$\kappa$-symmetry has been used to gauge-fix to zero
the eight fermionic parameters associated with $q_{aj}$.
Then the left-invariant currents $g^{-1}\p g$ now take the form
\eqn\bosm{J_{K^m} = [e^{-B}(\p \xt_n K^n +\p\tt_{aj} s^{aj}) e^B]_{K^m},
\quad
J_{s^{aj}} = [e^{-B} (\p \xt_m K^m +\p\tt_{bk} s^{bk}) e^B]_{s^{aj}}, }
$$
J_D = [e^{-B}\p e^B]_D, \quad J_{R_t} = [e^{-B}\p e^B]_{R_t} ,\quad
J_{\bar q_\ad^j} =
[e^{-B} \p e^B]_{\bar q_\ad^j},\quad
J_{\bar s^\ad_j} =
[e^{-B} \p e^B]_{\bar s^\ad_j},$$
$$J_{P_m}=0,
\quad J_{q_{aj}} = 0.$$
So the T-dualized action of \gsdualtwo\ reproduces the action of \gsactionnew\
if one uses the parameterization of \paramnew.

Finally, one can relate the parameterization of \paramnew\ with
the original parameterization of \gsuperp\ by using the isomorphism
of $PSU(2,2|4)$ which switches
\eqn\switches{P_m \to K^m, \quad q_{aj} \to s^{aj}, \quad \bar q_\ad^j \to
\bar s^\ad_j, \quad D \to -D.}
If one simultaneously switches the variables
\eqn\switchtwo{x^m\to \xt_m,\quad \t^{aj}\to \tt_{aj},\quad
\tb^\ad_j\to \xb_\ad^j, \quad y_t \to {y_t\over{|y|^2}},}
the parameterization of \paramnew\ is mapped to the parameterization
of \gsuperp. So it has been verified that after partially
gauge-fixing the $\k$-symmetry, the Green-Schwarz version of the $AdS_5\times S^5$
sigma model is mapped to itself under the combined $T$-duality with
respect to $x^m$ and $\t^{aj}$.

Since the argument above might have been too detailed, let us repeat the gist of the argument
using $SU(2|4)$ invariant notation. We group the coordinates as
$Z^{aI} = ( x^{a \dot a } , \theta^{a j} ) $ where $I$ is an $SU(2|4)$ index. We also
have the corresponding generators $G_{a I} = ( P_{a \dot a} , q_{a j})$ and their
dual generators $G^{a I} = ( K^{a \dot a} , s^{a j} )$.
We can then write the part of the action depending on $Z^{aj}$ as
\eqn\acti{
S \sim \int \epsilon_{ab} \eta_{IJ} M^{I}_{~ L} M^J_{~K} \partial Z^{a I} \bar \partial Z^{b K}
}
where $\eta_{IJ}$ is the supergroup invariant metric and $M^I_{~L}$ is given by
\eqn\formofm{
 Tr[ G^{a I} e^{- B} G_{b L} e^B] = \delta^a_b  M^I_{~L} ~,~~~~~~~
 Tr[ G^{aI} G_{b L}] = \delta^a_b \delta^I_L
}
After the T-duality we end up with dual variables $\widetilde Z_{aj }$ and the action will
be of a similar form but it will involve the inverse of this matrix. This inverse can
be written by an expression similar to \formofm\ but involving the inverse transformation
\eqn\forminv{
\delta^a_b  ( M^{-1} )^I_{~L} = Tr[ G^{a I } e^B G_{b L } e^{- B} ]
=  Tr[e^{-B} G^{a I} e^{B} G_{b L}]
}
where in the last expression we noticed that the inverse matrix can be viewed as the same
transformation $e^B$ as in \formofm\
but acting on the dual generators $G^{aJ}$.
Thus, after performing the transformation that exchanges the dual generators with the original ones
we end up with the same form of the action.

At the end of the previous subsection, we discussed an alternative
choice of T-dualization which also leaves the $AdS_5\times S^5$
background invariant. One can show invariance of the Green-Schwarz
sigma model using this alternative T-dualization by replacing the above
$SU(2|4)$ subgroup of $PSU(2,2|4)$ with the $PS(U(2|2)\times U(2|2))$
subgroup of \denomg. After gauge-fixing to zero the 8 fermionic
parameters associated with $s_a^{r'}$ and $\bar s_r^\ad$, one can
follow the same steps as above. One first groups the coordinates
as $Z_I^{J'} = (x_a^\ad, u_r^{r'}, \t_a^{r'}, \tb_r^\ad)$ where
$I=(a,r)$ and $J'= (\ad,r')$ are
$U(2|2)\times U(2|2)$ indices, and
$u_r^{r'}$ are four coordinates on the (analytic continuation of) $S^5$.
The corresponding generators are $G^I_{J'} = (P^a_\ad, R^r_{r'}, q^a_{r'},
\bar q^r_\ad)$ and their dual generators are
$G_I^{J'} = (K_a^\ad, R_r^{r'}, s_a^{r'},
\bar s_r^\ad)$. One can now repeat the procedures of \acti\ - \forminv\
to show that the action is mapped to itself under this T-duality.

\subsec{ Invariance of the $AdS_5\times S^5$ pure spinor sigma model}

In the previous subsection, it was shown that in the gauge
$\xi_{aj}=0$, the Green-Schwarz version of the $AdS_5\times S^5$ action
is invariant under T-duality where
$\xi_{aj}$ correspond to the 8 fermionic parameters
asssociated with the chiral superconformal generators $s^{aj}$.
In other words, the general element of the
${{PSU(2,2|4)}\over{SO(4,1)\times SO(5)}}$
coset is
\eqn\gene{\tilde g = g(x,y,\t,\bar\t,\bar\xi) \exp (\xi_{aj} s^{aj})}
where $g(x,y,\t,\bar\t,\bar\xi)$
is the gauge-fixed supercoset used in \gsuperp.
It will now be shown that the pure spinor version
of the $AdS_5\times S^5$ action is also invariant under T-duality.
Since the pure spinor version of the action is quantizable,
this proves that the sigma model action in an
$AdS_5\times S^5$ background is invariant under
T-duality to all orders in $\a'$.

The first step is to use the fact that there is a unique prescription
for constructing the pure spinor action from any $\k$-invariant Green-Schwarz action.
This prescription was first described by Oda and Tonin \oda\ and involves
relating the Green-Schwarz $\k$-transformations with the pure spinor BRST transformations.
So if the T-dualized Green-Schwarz action could be written in a $\k$-invariant form,
one could use this prescription to prove that T-dualization does not change
the pure spinor action.

However, the T-dualized Green-Schwarz action was only shown to be invariant
in the gauge $\xi_{aj}=0$. This means that the original and T-dualized
pure spinor actions may differ by terms which vanish when
$\xi_{aj}=0$. It will now be argued using BRST invariance that
such terms cannot be present.
Note that invariance under
T-duality of the BRST operators $Q=\int
dz \l^\a d_\a$ and $\hat Q=\int d\bar z\lh^\ah \hat d_\ah$
is manifest since the worldsheet variables
$(\l^\a,\lh^\ah)$ and $(d_\a,\hat d_\ah)$ transform by local
$SO(4,1)\times SO(5)$ Lorentz rotations under T-duality.

Suppose that the original pure spinor action is $S_0$ and
the T-dualized pure spinor action is $S_1$ where
\eqn\orig{S_1 = S_0 + \int d^2 z ~\xi_{a j} V^{aj}}
for some $V^{aj}$.
Then BRST invariance of $S_0$ and $S_1$ implies that
\eqn\invnew{\int d^2 z ~(Q+\hat Q)(\xi_{aj} V^{aj})= 0.}
Furthermore, as explained in \lads, $Q$ and $\hat Q$ act
on the supercoset element $\tilde g$ of \gene\ by right multiplication as
\eqn\rightmul{(Q +\hat Q) \tilde g = \tilde g [(\l^{aj} +\lh^{aj}) q_{aj} +
(\l_{aj} -\lh_{aj}) s^{aj} +
(\l^\ad_j -\lh^\ad_j) q_\ad^j +
(\l_{\ad}^{ j} + \lh_{\ad }^{j}) s^{\ad }_{j})]}
where the $j$ indices on $\l^{aj}$ and $\lh^{aj}$ are $SO(5)$
spinor indices which can be raised
and lowered using $(\s^6)_{jk}$ and $(\s^6)^{jk}$.
Using \gene\ and \rightmul, one learns that the only worldsheet
field which transforms into
$(\l_{aj}-\lh_{aj})$ is $\xi_{aj}$ which has the BRST transformation
$(Q+\hat Q)\xi_{aj} = (\l_{aj} -\lh_{aj}) + ... $
where the terms in $...$ will not concern us.

Suppose one expands
\eqn\expands{V^{a_1 j_1} = V^{a_1 j_1}_{(1)} + \xi_{a_2 j_2} V^{a_1 j_1 ~a_2 j_2}_{(2)} +
\xi_{a_2 j_2}\xi_{a_3 j_3} V^{a_1 j_1 ~a_2 j_2~a_3 j_3}_{(3)} +
 ...}
where $V^{a_1 j_1 ... a_n j_n}_{(n)}$ is assumed to be independent of $\xi_{bk}$ and
is antisymmetric under exchange of $a_k j_k$ and $a_l j_l$ indices.
Then if one focuses on terms in $(Q+\hat Q) (\xi_{aj} V^{aj})$
which are proportional to $(\l-\lh)_{aj}$ and have no $\xi_{aj}$ dependence,
\invnew\ implies that
\eqn\implex{(\l -\lh)_{a_1 j_1} V^{a_1 j_1}_{(1)} =0.}
Furthermore, since
$V^{a_1 j_1}$ can only depend on $(\l-\lh)_{aj}$ in
the ghost-number zero combinations of the Lorentz currents
$\l\g^{cd}w$ and $\lh\g^{cd}\hat w$, it is not difficult to
show that \implex\ implies that $V^{a_1 j_1}_{(1)}=0$.

One can then focus on terms
in $(Q+\hat Q) (\xi_{aj} V^{aj})$
which are proportional to $(\l-\lh)_{aj}$ and are linear in $\xi_{aj}$,
and use a similar argument to prove that $V^{a_1 j_1~ a_2 j_2}_{(2)}=0$.
Continuing to higher powers in $\xi_{aj}$, one proves that $V^{aj}=0$ and
therefore $S_0=S_1$ in \orig.

So it has been proven that
T-duality invariance of the $\k$ gauge-fixed Green-Schwarz action implies
that the pure spinor version of the action is also invariant
under T-duality.

\newsec{ Amplitudes and Wilson loops }

\subsec{Generalities on the  amplitudes}

In order to describe the external Yang Mills states it is convenient to use
an on-shell superspace formalism where the superfields $\Phi(x,\theta)$
depend only on the eight chiral superspace variables $\theta^{ai}$.
We also find it convenient to write four dimensional on-shell momentum as
\eqn\momp{
k_{a\dot a} = \pi_a \bar \pi_{\dot a }
}
which obeys $k^2=0$.
An on-shell gluon supermultiplet is characterized by a momentum $k$ and fermionic variables
$\kappa_i$ such that \refs{\FerberQX,\Nair}
\eqn\superf{
\Phi_{k,\kappa}(x, \theta) = e^{ i k \cdot x }
e^{  \pi_a \theta^{aj} \kappa_j }
}
Different components of the supermultiplet correspond to different terms in the $\kappa$
expansion. The $+$ helicity gluons correspond to the $\kappa^0$ terms and the $-$ helicity
gluons correspond to the $\kappa^4$ component.

The corresponding vertex operators in string theory have the form
\eqn\vertexop{
V_{\pi ,\bar \pi , \kappa}  =
 e^{ i  \pi_a \bar \pi_{\dot a } x^{a \dot a}  } e^{  \pi_b \theta^{bj} \kappa_j } \hat V
}
 where $\hat V$ can contain only derivatives
of $\theta$ and $x$. Of course, in addition it could contain other variables, such as $\bar\theta$,
with or without derivatives.
Thus the whole dependence on the $x$ and $\theta$ zero modes of the vertex operators comes
from the prefactor in \vertexop .

As we remarked above, before doing T-duality we should integrate out the zero modes of
$x^{a\dot a}$ and $\theta^{aj}$.
This implies that the amplitude   contains a factor
\eqn\factco{
{ \cal A}  =  \delta^4\left(\sum_{l=1}^n k^l_{a\dot a} \right) \, \,
\delta^8\left( \sum_{l=1}^n \pi^l_{a}  \kappa^l_i \right) \,\,  \widetilde { \cal A}
}
We can extract physical amplitudes for individual polarization states
from \factco\ by integrating over $\kappa^l$. Thus, if we simply integrate over $\kappa^l$ we would
be picking out the $(\kappa^l)^4$ term which is the minus helicity gluons. If we multiply by
$(\kappa^l)^4 $ and then integrate, then the $l$th particle corresponds to a + helicity gluon. This is
equivalent to setting $\kappa^l =0$ in \factco .
The presence of the fermionic delta function in \factco\
implies that the all  $+$ amplitude and the  almost all  $+$ and one $-$ amplitude
 vanish.  The first non-vanishing case is the MHV amplitude with mostly $+$ and two $-$
helicity gluons. For MHV amplitudes we do not need  any further $\kappa$ dependence in
$\widetilde{ \cal A}$, but amplitudes with more $-$ helicities
will require that we know the dependence
of $\widetilde {\cal A}$ on $\kappa$.  (A prescription for computing $\widetilde{\cal A} |_{\kappa =0}$
at tree level in string theory in flat space is given in appendix A.)

We   introduce an infrared regularization as follows.
 We imagine starting from a $U(N+k)$ theory. We consider a vacuum breaking the
 symmetry to $U(N) \times U(k)$ by giving a scalar field a vacuum expectation values
  $\mu_{IR}$ which  will play the
 role of an infrared cutoff. When we take the 't Hooft limit we   keep $k$ fixed, so that
 the low energy $U(k)$ theory becomes free. We then scatter $n$ gluons of the $U(k)$ theory.
 We  are  interested in the regime where all the kinematic invariants are much larger than
 the infrared scale, $s_{ij} \gg \mu_{IR}^2$. On the strong coupling side, this infrared
 regularization corresponds to introducing $k$ D3 branes in $AdS_5 \times S^5$. In terms
 of the $AdS$ metric $ds^2 = { dx^2 + dy^2 \over y^2}$ the branes are sitting at $y = 1/\mu_{IR}$.
 See figure 2. It is conceptually simpler for our purposes to say that $k=n$ and that the
 $n$ gluons are open strings that stretch among these $n$ branes so that each  portion of
 the
 boundary of the disk diagram corresponds to each  of the $n$ branes.

  \subsec{ Amplitudes after T-duality }

\ifig\tdualitymap{  The amplitude computation in the original theory involves the scattering
of open strings on $n$ D3 branes living in $AdS_5$. Under T-duality this maps to a different
computation in the T-dual $AdS$ space. The T-dual computation involves strings stretching between
$n$ D(-1) branes. The D(-1) branes are positioned so that the open strings between them are massless.
We are computing the interaction amplitude between these states in string theory which comes from
a disk diagram.
  } {\epsfxsize3.5in\epsfbox{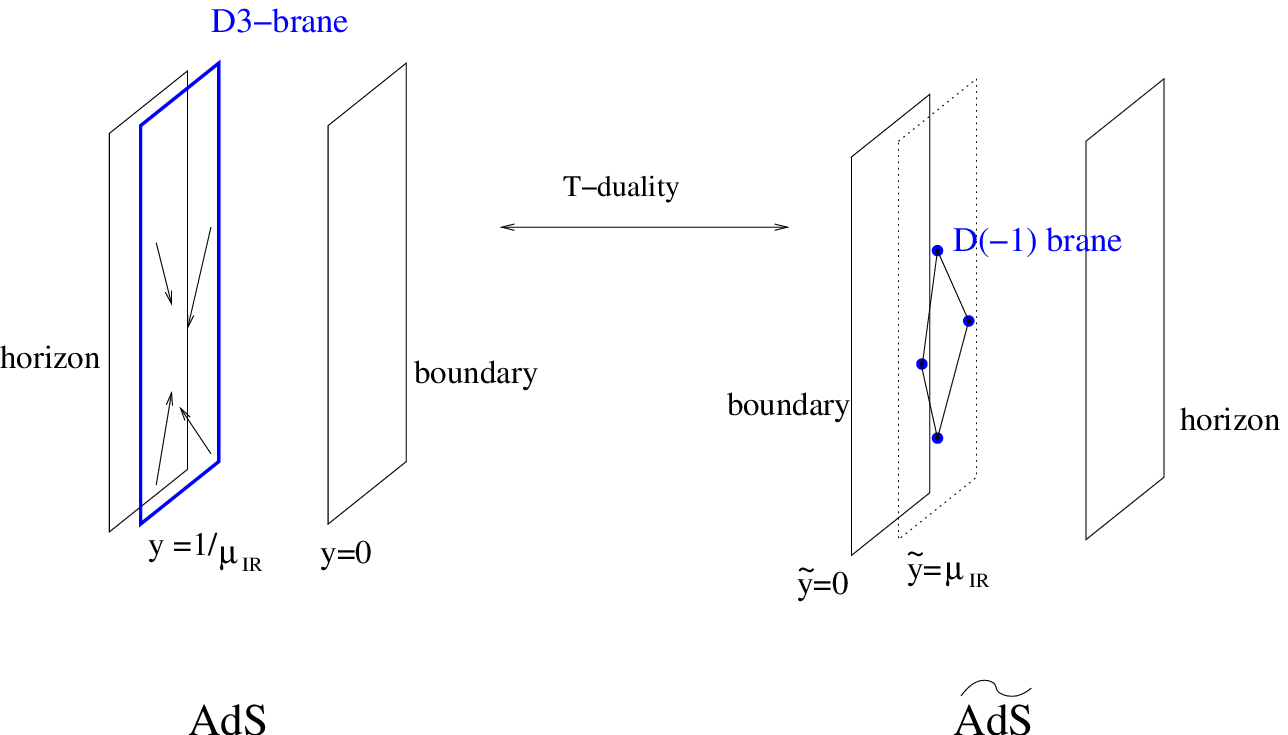}}

After T-duality we can compute the quantity $\widetilde {\cal A}$ in the T-dual theory. We explain
below what the corresponding computation is.
The T-dual computation of $\widetilde{\cal A}$
involves a number of D(-1) branes and each external state maps
to an open string stretching between the D(-1) branes, see \tdualitymap . All the $D(-1)$ branes
are sitting at the same $\widetilde y$ position $\widetilde y = \mu_{IR}$.
We can see that the open strings are stretched by looking at the original worldsheet
equation of motion
for one of the $R^4$ bosonic directions near
 the insertion point of the vertex operator \vertexop . It has the schematic form
\eqn\equmot{
 k_1 \delta^2(z) +  \partial  [  g_{11 } \bar \partial x^1 + \cdots ] +
  \bar \partial  [  g_{11 }  \partial x^1 + \cdots ] = k_1 \delta^2(z) +
  ( \partial \bar \partial - \bar \partial \partial) \widetilde x_1
 }
  where we have rewritten this equation in terms of the T-dual variable $\widetilde{ x}^1$. Integrating
  this in an arc around the insertion of the vertex operator at the boundary we conclude that
  $\tilde x^1$ has winding given by $k^1$. In other words, the boundary condition for $\widetilde{x}^1$ changes
  from one side of the vertex operator to the other by an amount proportional to $k_1$
Of course this is the familiar statement that momentum is mapped into winding under T-duality.
Let us now repeat this for the fermionic coordinates $\theta^{aj}$. We find that
the equation of motion is
\eqn\eqmotf{
 \pi_a \kappa_i \delta^2(z) + \partial  [  C_{a i \, b j }   \bar \partial \theta^{b j}  + \cdots ] -
  \bar \partial  [   C_{a i \, b j }   \partial \theta^{b j} + \cdots ] =  \pi_a \kappa_i
   \delta^2(z) +
  ( \partial \bar \partial - \bar \partial \partial) \widetilde \theta_{ai}
}
Thus we see that the T-dual fermionic coordinate $\widetilde \theta_{a i}$ has ``winding''
$\Delta \widetilde \theta_{a i} = \pi_a \kappa_i$ when we go across the vertex operator
insertion. Thus we can assign to each D(-1) brane also a position in  $\widetilde \theta$ which
is consistent with these jumps. Notice that we will not integrate over the overall $\widetilde \theta$
fermion zero mode, so we are allowed to fix the position of one of these D(-1) branes arbitrarily.
The same is true for the bosonic zero modes. One of the $D(-1)$ brane positions is fixed
arbitrarily.
We have $n$ D(-1) branes, at specific separations given by the momenta and the fermionic
coordinates $\kappa_i$ of the external gluons.
We have open strings stretching  between them that are on-shell.
Then we compute a disk diagram which is the tree level contribution to the interaction
between these open strings. The whole computation is done in terms of the T-dual model, which
is T-dual conformal invariant. The information about the polarizations of the gluons appears
as the information about the particular open string state stretching between D(-1) branes
that we are considering and it is encoded by the $\kappa$ variables.
We see that the theory, written in terms of the T-dual variables, has
manifest dual superconformal symmetry, up to a small subtlety.
If we consider the regularized amplitude, with a finite $\mu_{IR}$, as in \tdualitymap , then
we map this to a configuration of
  D(-1) branes at the same position of the radial variable
of the T-dual AdS space, $\tilde y = \mu_{IR}$.
 However, a dual special conformal transformation will change their relative radial positions.
 In the limit that $\mu_{IR} \to 0$, these positions are formally all
at $\tilde y=0$, which is the boundary of $\widetilde{AdS}$ space, and are unchanged by the
conformal transformation. However,  the action
diverges. Fortunately the structure of the divergences is known. 
After   extracting the IR divergencies, one
finds that the amplitude changes in a well defined way under such conformal transformation.
The change is completely fixed by the structure of the IR divergencies.
This was discussed in detail in  \sokatchevward\ (see also \aldaylectures\ for a string perspective
on the same issue.).

The bottom line is that the T-duality argument makes manifest the T-dual conformal symmetry
and explains why it should be a symmetry of the amplitude.  We have not been very
explicit about the precise form of the vertex operators, but it seems clear
that the symmetries are such that one should reproduce the structure described in \nmhv\ (and also
  \sokatch ).

\subsec{ The amplitude and the Wilson loop }

Let us now turn to the Wilson loop computation. The Wilson loop computation involves
a string configuration very similar to the one that we get after performing the T-duality and
taking $\mu_{IR} \to 0$.
One difference is that in the Wilson line computation there is no information about the
polarization states of the gluons. This information arises in the T-dual computation
as the polarization information for the strings stretching between $D(-1)$ branes.
In order to obtain the Wilson loop, we need to ``forget'' about these polarization states
and reduce the computation to one with fixed boundary conditions on the boundary of the string.
For example we will put Dirichlet boundary conditions for the fermions and also for
the $AdS$ bosons.
In the particular case of MHV amplitudes we expect that this change will simply produce
a factor proportional to the tree level MHV amplitude.
In other words, on the basis of the perturbative
computations done in \refs{\DrummondAUA,\brandhuber,\sixgluon,\sixwilson}, we expect that
the relation is
\eqn\relatio{
\left. \widetilde { \cal A} \,  \right|_{\kappa =0}  =
{ 1\over \prod_{i=1}^n (\pi_i \pi_{i+1} ) } \langle W(k_1, \cdots ,k_n) \rangle
}
up to IR and UV divergent terms.
We do not have a rigorous
justification for the origin of this prefactor on the string theory side.
Of course,  this factor accounts properly for the right helicity weights of the amplitude.
  It was also argued in \nmhv\ that it is dual superconformal covariant with weights one.
   So the only issue is whether one could get a residual superconformal
invariant factor.

In lieu of a derivation, let us give some plausibility arguments.
From the field theory side, as pointed out in \mcgreevy , when we put in
this regularization we
have an outer loop in the Feynmann diagram which consists of  a massive supermultiplet.
This particle mediates the interaction between the external $U(k)$ particles
 and the $U(N)$  internal particles.
In the limit that we turn off the Yang Mills coupling of the U(N) theory, then we simply
have a one loop diagram and we know that the result is equal to the MHV tree amplitude up to
terms that capture the IR divergences \BernQK  .
From the string theory side, it is clear that the only difference between the amplitude and the
Wilson loop computation lies in the detailed boundary conditions at the boundary of the worldsheet.
(We explore the shape of the worldsheet near the boundary for a finite $\mu_{IR}$ in appendix B.)
Thus, it seems natural to expect that
 the worldsheet theory will contain
some worldsheet excitations that are confined to the boundary of the worldsheet that would
give rise to the prefactor in \relatio . These could
represent the massive particle we have in the field theory traveling around
the loop. Then the difference between the amplitude and the Wilson loop would be whether we
do or do not include these degrees of freedom localized at the boundary.

It is also quite plausible that we need to consider a Wilson loop with some insertions
that take into account the polarization states of the particles.
Studying the string theory in more detail one should  be able to give a definite answer
to these questions. It is also possible that one could understand this prefactor by
computing precisely, as in \mcgreevy , the relation between MHV amplitudes and momentum
space Wilson loops.

\newsec{ Dual conformal symmetry in the AdS sigma model}

In this section we consider  a bosonic sigma model with an
$AdS_{d+1}$ target space. Our goal is to get some insight on the connection
between the dual conformal symmetry and integrability. The conserved currents
associated to integrability in the original and T-dual model were studied in \tseytlintdual\ and
the flat connection of the T-dual model was written in terms of the variables of the original
model. Our goal here is closely related.  We will first
relate the non-trivial dual conformal generators with the non-local
currents that arise through integrability. We will also show the gauge equivalence of
the flat connection of the original model and the one arising from the T-dual model.

As we mentioned above,
writing the metric as \eqn\metrb{ ds^2 = { dx^2 + dy^2 \over
y^2} } and performing T-duality in the $x$ coordinates and an
inversion of $y$
\eqn\dualc{
 d \widetilde x = * { d x \over y^2 } ~,~~~~~~~~ \widetilde y = { 1 \over y }   ~,~~~~~~~~
  d  x = * { d \widetilde  x \over {\tilde y}^2 }
 }
we can see that the equations of motion for $\widetilde x $ and $\widetilde y$
are the same as the ones we would obtain for a sigma model on the
T-dual AdS space, or $ \widetilde{ AdS}_{d+1} $ space $ds^2 = { d {\tilde x}^2 + d {\tilde y}^2 \over
\tilde y^2 } $. The new AdS space has an $SO(2,d)$ symmetry group.
Some of these symmetries are the same as the symmetries of the original model.
For example, the dilatation symmetry $D$ of the original model is related
to the dilatation symmetry of the dual model, $D = - \widetilde D$. On the other hand,
the special conformal symmetries of the dual model   are not
so obvious in the original model.
We would like to understand what these symmetries are in the original model.

Let us consider first the simpler example of Euclidean $AdS_2$ or $H_2$.
 In this case
we can write the special conformal generator of the dual model as
\eqn\specconf{
\widetilde K = \int d \sigma j^{ \tilde K}_\tau(\sigma) = \int d\sigma \left[ (\tilde x^2 -\tilde y^2)
{ \partial_\tau \tilde x \over \tilde y^2 }
 + 2 { \tilde x \partial_\tau \tilde y \over \tilde y} \right]
}
where $\tau$ and $\sigma$ are the time and space coordinates on the worldsheet.
We can now use \dualc\ to replace the time
 derivatives of $\tilde x$ by sigma derivatives of $x$. We can then integrate these
 by parts. In this integral there
 are boundary terms and we assume that we can ignore these boundary terms (this will
  be true in the application we
 have in mind, where we will integrate on a closed contour and  
 demand that $x$ and $\tilde x$ are periodic). We are left with terms of the form $ \tilde x
 \partial_\sigma \tilde x  x $.
 We now write $\tilde x(\sigma) = \int^\sigma d\sigma' \partial_\sigma \tilde x $, and  we
 replace the derivatives of $
 \tilde x$ by derivatives of $x$ using \dualc\ again.
 In the end we are left with an expression of the form
 \eqn\specfs{
 \tilde K = \int d\sigma \int^\sigma d \sigma'  j^P_\tau (\sigma') j^D_\tau(\sigma)  +
  \int d \sigma j^P_\sigma
= P_2}
where $j^P \sim { \partial x \over y^2 }$,  $j^D \sim { x d x + y d y \over y^2 }$  are
the the translation and dilatation currents  of the original model.
Thus we see that the special conformal transformation in the dual model correspond
 to one of the non-local
conserved charges. It is the second non-local conserved charge
which is given by two integrals.
 Since the AdS model is integrable, we have an infinite set of non-local charges.

Thus, the conclusion is that the conformal symmetry of the dual model maps to the
 higher non-local charges of
integrability. The same result is true in general $AdS_{d+1}$ spaces. Thus, when we demand
that a certain quantity is invariant under the dual conformal symmetry we are demanding that
it is invariant under some of the non-local charges associated to integrability.

A simple way to think about these non-local charges is to construct a one
parameter family of flat connections
${\cal C}(\lambda)$. This one parameter family can be used to write all
the non-local conserved charges
as we will review below.
We can do the same for the dual model and construct ${\widetilde{\cal C}}(\lambda) $.
We will then show
that these two connections differ only by a gauge transformation,
so that the total set of  charges is the  same on
both sides.

\subsec{Integrability and the flat connection}

We work in $AdS_{d+1}$. This is described in terms of the coset manifold $SO(2,d)/SO(1,d)$ or $G/H$.
We think of it as a right coset $g \sim g h $.
  The group $G$ acts on the left and it corresponds to   the global isometries of $AdS$.
We will now construct the conserved currents for the model, following
 the discussion in \polch (see also \MandalFS ), with some minor changes in notation.
We construct the left invariant ($G$ invariant) currents
\eqn\curret{
J = - g^{-1} d g
}
and we decompose them according to the decomposition of the Lie algebra
${ \cal G} = {\cal H} + { \cal M}$, where ${\cal H}$ are the generators in the subgroup $H$
and ${\cal M}$ are the rest. We then find
\eqn\decomp{
J = H + M
}
This transforms under ${\cal H}$ gauge transformations.
The quantity
\eqn\quant{
m = g M g^{-1}
}
is ${\cal H}$-invariant. The lagrangian can be written as $L \sim Tr[ m_\alpha m^\alpha] \sim
Tr[ M_\alpha M^\alpha ] $.
For two quantities related as $x = gX g^{-1} $ we will use the lower case letter for the $H$
invariant version and the upper case letter $X$ for the G invariant one.
We also note that
\eqn\vers{
dx = g dX g^{-1} - j \wedge x - x \wedge j
}
where $j$ corresponds to $J$.
Since  $H$ is a subgroup we have $[{\cal H} , {\cal H} ] \subset {\cal H}$. Since we are
performing a coset we also see that
  $[{\cal H} , {\cal M} ] \subset {\cal M}$. In our case we also have
 $[{\cal M} , {\cal M} ] \subset {\cal H}$  \foot{ This can be seen as follows. Up to an irrelevant
 change in signature the coset is the same as $SO(1,d+1)/SO(d+1)$. Then ${\cal H}$ are all the rotation
 generators
 and ${\cal M}$ are all the boost generators.
  We know that the commutator of two boost generators is a rotation. }.
 From the definition \curret\ we know that $dJ= J\wedge J$.
 Decomposing $J$ as in \decomp\ and equating both sides we get
 \eqn\decomps{
 d H = H \wedge H + M \wedge M ~,~~~~~~~~~~~~d M = H \wedge M + M \wedge H
 }
 This then implies that
 \eqn\versim{
 d m = - 2 m \wedge m
 }

 In addition we also have that $m$ is proportional to the Noether current for the left $G$ action.
 So $ d * m =0$.
 Thus we construct the flat connection as
 \eqn\intecunt{
 {\cal C } =- 2 \sinh^2 { \lambda \over 2 } \, \,  m +
  \sinh { \lambda } \,  *  m
 }
 where $\lambda$ is an arbitrary complex parameter. This obeys $ d {\cal C} + { \cal C} \wedge
 {\cal C} =0$.
  One can then construct the holonomy
  \eqn\omedefi{
  \Omega(\lambda) = P e^{\int {\cal C(\lambda) }}
  }
  Expanding this in powers of $\lambda$ we get an infinite set of non-local conserved charges.
  The charge $Q_n$ multiplying $\lambda^n$ will contain a maximum of $n$ integrals.

  In the case of the cylinder we  need to consider
  $Tr[ \Omega^n]$ where $\Omega$ is the holonomy around the cylinder. These are then
  the conserved charges for a cylinder.


  In the application to the amplitude we have a worldsheet which is a disk and
   thus we can form the holonomy around the origin of the disk. Since this can be
   smoothly deformed to the origin we conclude that the holonomy should be simply the
   identity matrix $\Omega =1$.
   This is stating that the amplitude should be annihilated by all the charges, both the
   local and non-local charges\foot{This point of view was emphasized to us by A. Polyakov and
   A. Murugan.}. We see that dual conformal symmetry corresponds to the statement that
   some particular charges annihilate the amplitude. Of course one needs to treat IR divergences
   carefully (see \sokatchevward ),
    but this is the essence of the statement.  It is natural to expect that demanding
   that all non-local symmetries annihilate the amplitude should determine the amplitude.

  \subsec{ Relation between the flat connection in the original and the T-dual model}

 We now make a specific choice for the coset representative $g$ as
 \eqn\formofg{
 g = e^{ x. P } e^{ \log y D }
 }
 where $D$ is the dilatation operator and $P_i$ are the momenta, $i=1,\cdots ,d$.
 We have $[D,P_i] = P_i$.  We also have the special conformal generators
 $K_j$, $[D,K_j] = - K_j$, $[ K_i , P_j] = 2 \delta_{ij} D + $rotation .
 Note that a combination of $P$ and $K$, $\half ( P + K)$ is in $H=SO(1,d)$ while the other
 combination is not.
 We have that
 \eqn\decomp{ \eqalign{
 J = &  - \left[ { dy \over y } D + { dx^i \over y } P_i \right] =
  - \left[ { dy \over y } D + { dx^i \over y } \half ( P_i - K_i)  \right] - { dx^i \over y}
  \half ( P_i + K_i)
  \cr
  M = & -  \left[ { dy \over y } D + { dx^i \over y } \half ( P_i - K_i)  \right]
  \cr
  H = &  - { dx^i \over y}
  \half ( P_i + K_i)
  }}
We can now construct the flat connection as in \intecunt .
  It is now convenient to do a gauge transformation of ${\cal C} \to
  {\cal C}' = g^{-1} {\cal C} g +
  g^{-1} d g $, where $g$ is given in \formofg .
  We then get
  \eqn\versi{\eqalign{
  {\cal C}' = &  - 2 \sinh^2 { \lambda \over 2 } M +  \sinh { \lambda }
   *   M - ( H + M) = - \cosh \lambda M + \sinh \lambda * M - H
  \cr
  {\cal C}' = & ( \cosh \lambda { dy\over y} - \sinh \lambda * {dy \over y} ) D
  +
  \cr
  &+ \cosh { \lambda \over 2 } ( \cosh{ \lambda \over 2 } { d x^i \over y } - \sinh{ \lambda \over 2 }
  * { dx^i \over y } ) P_i +
  \sinh {\lambda \over 2 } ( - \sinh {\lambda \over 2 } { d x^i \over y } + \cosh {\lambda \over 2 }
  * { d x^i \over y } ) K_i
  }}

  We can now construct a similar current in the $T$ dual model, $\widetilde {\cal C}$, and then
  make a similar gauge transformation but in the T-dual model. We then get
  \eqn\currenttd{ \eqalign{
   \widetilde {\cal C}' = &
   ( \cosh \lambda { d\tilde y\over \tilde y} - \sinh \lambda * {d\tilde y \over \tilde y} ) D
  +
  \cr
  &+ \cosh { \lambda \over 2 } ( \cosh{ \lambda \over 2 } { d\tilde x^i \over \tilde y } -
   \sinh{ \lambda \over 2 }
  * { d\tilde x^i \over \tilde y } ) P_i +
  \sinh {\lambda \over 2 } ( - \sinh {\lambda \over 2 } { d \tilde x^i \over \tilde y } +
   \cosh {\lambda \over 2 }
  * { d \tilde x^i \over \tilde y } ) K_i
  }}
  In principle we could have introduced another parameter
   $\tilde \lambda$ here. But, anticipating
  our result, we have set $\tilde \lambda = \lambda$.
  We can now express $\tilde {\cal C}'$ in terms of the original variables ($x$ and $y$) via
  \dualc . We then make an additional gauge transformation of ${\cal C}'$, this time by a
  constant group element,  which
    maps
  $D \to - D$ and $P \leftrightarrow K$.

We then find \eqn\currenttd{ \eqalign{
   \tilde {\cal C}'' = &
   ( \cosh \lambda { d  y\over  y} - \sinh \lambda * {d  y \over   y} ) D
  +
  \cr
  &+ \cosh { \lambda \over 2 } ( \cosh{ \lambda \over 2 }  * { d x^i \over  y } -
   \sinh{ \lambda \over 2 }
    { d  x^i \over  y } ) K_i +
  \sinh {\lambda \over 2 } ( - \sinh {\lambda \over 2 } * { d   x^i \over   y } +
   \cosh {\lambda \over 2 }
  { d  x^i \over   y } ) P_i
  }}

  We now note that the original flat connection $ {\cal C}'$ can be
  related to $\tilde {\cal C}''$ via a gauge transformation by a constant group element
  \eqn\constgaugt{
  {\cal C}' = e^{ - \mu D} \tilde {\cal C}'' e^{\mu D }
  }
  where $\mu$ is given by
  \eqn\defmu{
  e^\mu = \tanh { \lambda \over 2 } ~,~~~~~~~e^{ - \mu D} P e^{\mu
  D} = e^{ -\mu } P ~,~~~~~~~~e^{ - \mu D} K e^{\mu
  D} = e^{ \mu } K
}

  We can see that expanding \constgaugt\ in powers of $\lambda$ one
obtains a relation between non-local currents of different order. Notice that
  the gauge transformations we used prior to \constgaugt\ were  $\lambda$ independent.

We have recently learnt that similar results, including a generalization to
the full  $AdS_5 \times S^5$ coset theory were obtained in \TseytlinNew .

\newsec{ Conclusions}

In this paper we have discussed the concept of ``fermionic T-duality''. We have
shown that this is a symmetry of tree level string theory. At the level of
the worldsheet we are performing the same steps as the ones we perform for a
bosonic T-duality. We select a fermionic variable $\theta$ which has a shift symmetry.
This corresponds to a supersymmetry that anticommutes to zero, $Q^2 =0$.
 We then introduce the dual variable $\tilde \theta$ via equations that are similar
 to the ones we use for a bosonic T-duality.  In target space this maps one supersymmetric
 background to another supersymmetric background. The RR fields and the dilaton are changed
 but the metric and the $B$ field remain the same. In general, the reality conditions are not
 respected because we need a complex Killing spinor in order to have
 $Q^2=0$ for the corresponding supercharge. If we restrict to fermionic variables which
 are single-valued on the worldsheet, the T-duality will probably not extend to higher orders in string
 loops. On the other hand we expect it to be exact in $\alpha'$.
 In fact, the change  of the dilaton comes from a determinant that appears
 when we perform the change of variables in the path integral, as in the bosonic case
 \refs{\BuscherQJ,\SchwarzTE}. One example is the case of constant graviphoton background.
 This results from performing fermionic T-duality on a flat space background after adding
 a total derivative term to the action. Thus tree level string theory on a constant graviphoton
 background is the same as string theory on flat space. At higher string loop
 orders the two are different.

  We have then applied this idea to the $AdS_5 \times S^5$ background. We performed four
  bosonic T-dualities along four translation symmetries of $AdS_5$ as well as eight fermionic
  T-dualities along the directions associated to the chiral Poincare supersymmetry generators
  $Q_{a i}$ where $a$ is a four dimensional chiral spinor index and $i$ is a fundamental $SU(4)$
  R-symmetry index. After the dualities, the string theory comes back to itself. But the initial
  problem of computing scattering amplitudes translates into a problem involving a certain
  D(-1) brane configuration that is very similar to a Wilson loop configuration, with the
  D(-1) branes at the corners of the Wilson loop.
  The ordinary superconformal symmetry of the dual superstring theory is what was called
  ``dual superconformal symmetry'' of the original theory. Thus, this transformation makes
  this dual symmetry manifest.
   We have argued
  this for the classical Green Schwarz sigma model and then for the full quantum theory
  constructed using the pure spinor formalism.
  Our arguments amount to a change of variables in the path integral. We expect that there
  should be no anomalies associated to it. In particular, at one loop, we have checked
  that the Jacobian for this change of variables vanishes.
  Thus, we expect that the symmetry should be a full symmetry for any value of $\lambda = g^2 N$.
  In other words, we expect it to be exact in $\alpha'$.
  This then explains the presence of the dual conformal symmetry found in weak coupling
  computations \refs{\sokatch,\sixgluon,\sixwilson,\nmhv}. It would also be nice to explain the emergence of this symmetry purely within
  the weak coupling theory. The four bosonic T-dualities are essentially a Fourier transform.
  This paper suggests that it would be productive to try to perform an additional transformation
  of the fermionic variables in order to be able to see  the duality.

  In the context of the simpler bosonic $AdS$ sigma model we have also shown that the
  dual conformal symmetry amounts to some subset of the non-local charges associated
  to integrability. This has also been recently been done, including the extension to the
  full $AdS_5 \times S^5$ sigma model,  in \TseytlinNew.

  It has become clear that ``dual superconformal symmetry'' is very powerful in restricting
  the form of the amplitude. It even fixes the full amplitudes for four and five gluons
  \sokatchevward .
  Since this symmetry is simply a small part in the infinite set of conserved charges associated
  to integrability one would hope that all the higher charges can similarly be put to use
  in order to fully fix all amplitudes.

  Fermionic T-dualities probably have many more applications that the one we used in this paper.
 In particular, since fermionic T-duality is a symmetry of supergravity, it seems that
 it might be possible to  consider the continuous symmetry groups (e.g. $E_7$ \Cremmer )
  that arise from
 toroidal compactifications and extend them to supergroups.
If the current discussion of $E_{10}$ and $E_{11}$
models (see \refs{\damour,\west} for recent papers)
could be generalized to supergroup models, it might be possible
to derive the d=10 and d=11 fermionic supergravity fields in
the same manner as the bosonic supergravity fields have
been derived in these models.

{ \bf Acknowledgements}

We thank Fernando Alday, Jacques Distler, James Drummond,
Chris Hull, Arvind Murugan,  Martin Ro\v cek, Warren Siegel,
Emery Sokatchev, Arkady Tseytlin,  Cumrun Vafa and E. Witten for discussions.
We also thank the organizers of the Paris workshop ``Wonders of Gauge
Theory and Supergravity'',  the Stony Brook workshop
``Simons Workshop in Mathematics and Physics 2008''
and the Mumbai ``Monsoon workshop in string theory''
where parts of this work were presented.
N.B. also thanks the IAS at Princeton where part of this work was done.
 We would also like to thank the authors of \TseytlinNew\ for sharing a draft of
 their paper with us, and for patiently waiting until we finished writing our
 paper to submit theirs.

The work of N.B. was partially supported by CNPq grant 300256/94-9
and FAPESP grant 04/11426-0.
The work of J.M. was supported by DOE Grant DE-FG02-90ER40542.

\appendix{A}{ MHV Tree Amplitudes in Superstring Theory}

MHV tree amplitudes in flat space
open superstring theory were studied in \stieberger\ using
the RNS prescription.
In this appendix, we propose a new prescription for computing MHV
tree amplitudes in open superstring theory. Although our original
motivation was to compute MHV superstring
tree amplitudes in an $AdS_5\times
S^5$ background, up to now we have only been able to develop this
prescription in a flat background. Nevertheless, this flat space
prescription
for computing superstring MHV tree amplitudes is simpler than
previous prescriptions and has an interesting relationship with
the self-dual N=2 string of \refs{\ademollo,\oogv}.
Such a relationship is not surprising since the self-dual
N=2 string computes self-dual d=4 Yang-Mills amplitudes which have
many features in common with MHV amplitudes \cangemi\chalmers\bern.

\subsec{MHV tree amplitudes in gauge theory}

$N$-point tree-level MHV amplitudes have an extremely simple form
when expressed in terms of spinor helicities. If the $d=4$
light-like momentum $p_{a\ad} = p_m \s^m_{a\ad}$ of the $r^{th}$
state is written as \eqn\twi{p_r^{a\ad} = \pi_r^a\pibar_r^\ad, }
the color-ordered $N$-point tree-level MHV amplitude with $N-2$
self-dual gluons and 2 anti-self-dual gluons is \eqn\amhv{A =
{ (\pi_J \pi_K)^4 \over \prod_{r=1}^N (\pi_r \pi_{r+1})}} where $J$
and $K$ label the anti-self-dual gluons, $\pi_{N+1}\equiv\pi_1$,
and the color factor $Tr(T^{a_1} ... T^{a_N})$ has been
suppressed. In \amhv, the self-dual gluon polarization is
$\eta_r^{a\ad}=\ve_r^a\pibar_r^\adot$ and the anti-self-dual gluon
polarization is $\bar\eta_r^{a\ad}=\pi_r^a\ve_r^\adot$ where
$\ve_r^a$ and $\ve_r^\ad$ are normalized such that
$\ve_r^a\pi_{ra}=1$ and $\ve_r^\ad\pibar_{r\ad}=1$.

The formula \amhv\ can be easily extended to describe the
scattering of any ${\cal N}=4$ super-Yang-Mills fields by
combining the $\cal N$=4 super-Yang-Mills fields into a scalar
chiral superfield $\Phi(x,\theta)$. For an on shell gluon the
field has a special form characterized by its momentum and some
fermionic parameters $\kappa^i$ determining its various
components. We have \eqn\expans{ \Phi_{p,\kappa} ( x , \theta) =
e^{ i p .x } e^{ \pi_a \kappa_i \theta^{a i} }
}
Expanding in powers of $\kappa$ we obtain the various components of the
superfield.
We can think about the amplitude as a function of $\pi^a , \bar \pi^{\dot a} , \kappa_i$ for
each gluon. By looking at the $(\kappa^r)^4$ term we extract the amplitude
for the negative helicity  gluon, while  the $(\kappa^r)^0$ term
corresponds to the  positive helicity gluon.
%

The amplitude will contain an integral over $\theta$ that will translate into an
overall factor of the form
\eqn\amplit{
{\cal A}(\pi^r , \bar \pi^r , \kappa^r) = \delta^4( \sum_r p_r)
 \delta^8 ( \sum_r \pi^a_r \kappa^r_i )
\widetilde { \cal A}
}
The amplitude $\widetilde {\cal A}$ could have additional $\kappa$ dependence.
However, the $\kappa$ independent part of $\widetilde {\cal A}$ is the MHV amplitude, up
to the prefactor in \amplit .

In field theory we find that this MHV part is given by
\eqn\supermhv{
\widetilde {\cal A} =  { 1 \over \prod_{r=1}^N (\pi_r \pi_{r+1})}
}
The numerator factor in \amhv\ comes from considering the $\delta$ function in \amplit\ and
integrating over four of the $\kappa$'s for each of the negative helicity gluons.


\subsec{MHV tree amplitude in open superstring theory}

The arguments leading to \amplit\ were completely kinematical and   also hold
for open superstring theory. Namely, they   also hold if we consider open string
scattering for massless open strings on a D3 brane,  even in the case that the
scattering occurs at energies higher than the string scale. In that case the
MHV amplitude will not be given by \supermhv\ and will contain dependence
on $\a'$ . In this subsection we propose a
way to compute $\widetilde { \cal A} $ in flat space open superstring theory.
%

We conjecture that the MHV superstring amplitude is given by
\eqn\stringa{ \eqalign{ \widetilde{ \cal A}(\pi_r , \bar \pi_r )   =&
(\pi_1 \pi_2)^{-1} (\pi_2 \pi_3)^{-1} (\pi_3 \pi_1)^{-1} \times
\cr
 & \langle V_1(z_1) V_2(z_2) V_3(z_3) \int_{z_3}^{z_1} dz_4 U_4(z_4) ...
\int_{z_{N-1}}^{z_1} dz_N U_N(z_N) \rangle
}}
where $V_r(z_r) =  e^{i\pi_r\pibar_r x(z_r)}$ and
\eqn\vertU{U_r(z_r) = (\ve_r^a \pibar_r^\ad \p x_{a\ad} +
\psi_\ad \bar\psi_\bd \pibar_r^\ad \pibar_r^\bd) e^{i\pi_r\pibar_r x(z_r)}.}
The correlation function in \stringa\ is defined in the usual manner
where $x_{a\ad}(z)$ satisfies the OPE $x_{a\ad}(y) x_{b\bd}(z) \to
-\a' \e_{ab}\e_{\ad\bd} (\log |y-z| + \log |y-\bar z|)$
and $(\psi_\ad,\bar\psi_\bd)$ are fermions of conformal weight
$(\half,0)$ satisfying the OPE
$\psi_\ad(y)\bar\psi_\bd(z)\to \a' \e_{\ad\bd} (y-z)^{-1}$.
If $(\psi_\ad,\bar\psi_\bd)$ are relabeled as $\psi_{a\ad}$, the vertex
operator $\int dz U(z)$ of \vertU\ is the standard RNS vertex operator
for a self-dual gluon.
Note that $U_r(z_r)$ changes by a total derivative under
the gauge transformation $\d \ve_r^a = c\pi_r^a$ for any constant $c$,
so with the normalization $\ve_r^a \pi_{ra} =1$, the amplitude
is independent of $\ve_r^a$.

The novelty of \stringa\ is that the computation of the
superstring MHV tree amplitude is manifestly invariant under
$N=4$ $d=4$ spacetime supersymmetry. Although one can of course
compute superstring MHV tree amplitudes using either the RNS or
pure spinor formalism, computations using these
formalisms are more complicated and contain many more fields.
Note that \stringa\ depends only on $\pi_r \bar \pi_r$ but not on $\kappa$.
This is because we are concentrating on MHV amplitudes and we are only giving
a prescription for computing MHV amplitudes. We are not saying how to compute
non-MHV amplitudes, which should contain some $\kappa$ dependence.

We will not attempt to derive \stringa\ from a superstring formalism,
however, there is an interesting relation to the open self-dual
string with N=2 worldsheet supersymmetry \refs{\ademollo,\oogv}.
This open string theory
has a single physical state in its spectrum corresponding to
a self-dual Yang-Mills gluon (in signature $d=(2,2)$).
The worldsheet matter variables in the self-dual string consists
of $(x^{a\ad}, \psi_\ad, \bar\psi_\ad)$ with $\hat c=2$ $N=2$
superconformal generators
\eqn\twosc{T = \half \p x^{a\ad}\p x_{a\ad} + \half
(\psi^\ad\p\bar\psi_\ad + \bar\psi^\ad\p\psi_\ad), \quad
G^+ = \psi_\ad \p x^{+\ad}, \quad
G^- = \bar\psi_\ad \p x^{-\ad}, \quad J= \psi^\ad \bar\psi_\ad,}
and worldsheet action
\eqn\twoact{{1\over{\a'}}\int d^2 z [ \half \p x^{a\ad} \pb x_{a\ad}
+ \bar\psi^\ad \pb \psi_\ad ].}

The physical self-dual Yang-Mills state is associated with the
$N=2$ superconformal primary field
\eqn\prim{V = \exp (i p_{a\ad} x^{a\ad}),}
and the integrated vertex operator is
\eqn\intvert{\int dz G^- G^+ V =
\int dz \pi^- (\bar\pi_\ad \p x^{+\ad} + \pi^+ (\psi_\ad \bar\pi^\ad)
(\bar\psi_\ad \bar\pi^\ad) ) e^{i p x}}
where $p_{a\ad} = \pi_a\bar\pi_\ad$.
So if one chooses the gauge $\e^+ =0$ and $\e^- = {1\over{\pi^+}}$,
\intvert\ is equal to $\pi^+\pi^- \int dz U(z)$ where $U(z)$
is defined in \vertU.

Using the ``topological'' rules of \vafatopo\ for computing
self-dual open string amplitudes, the $N$-point tree amplitude
prescription is
\eqn\presc{{\cal A}_{{\cal N}=2} = \langle (G^+ V(z_1))(G^+ V(z_2)) V(z_3)
\prod_{r=4}^N \int dz_r U_r(z_r) \rangle}
where the N=2 superconformal generators of \twosc\
have been twisted so that
$\psi_\ad$ carries zero conformal weight
and the zero-mode measure factor is $\langle \psi_\ad \psi^\ad \rangle=1$.
As shown in \vafatopo, these $N$-point amplitudes vanish when $N>3$
as is expected for self-dual Yang-Mills tree amplitudes.

The $N$-point tree amplitude prescription proposed here is slightly
different from \presc\ and is
\eqn\presc{ \widetilde{\cal A} = (\pi_1\pi_2)^{-1} (\pi_2\pi_3)^{-1}
(\pi_3\pi_1)^{-1}~~~ \langle  V(z_1)~ V(z_2)~ V(z_3)~
\prod_{r=4}^N \int dz_r U_r(z_r) \rangle}
where the N=2 superconformal generators are untwisted. As shown below,
this new prescription is non-vanishing for $N>3$ and reproduces
the gauge theory result of \supermhv\ in the limit
when $\a'\to 0$.

This suggests that  there should be a superstring formalism which
combines the worldsheet variables of the self-dual string with
another sector containing $\t^{aj}$ worldsheet variables. One possibility
for such a formalism is the self-dual super-Yang-Mills string
theory constructed in \symtwistor, which is related to the
Green-Schwarz self-dual string of \supersieg.
It would be very interesting if one could use this
formalism to construct a prescription with manifest
$N=4$ $d=4$ supersymmetry which
reproduces superstring non-MHV tree amplitudes.

\subsec{$\a'\to 0$ limit of superstring tree amplitude}

The first step in checking the validity of \stringa\
is to show that it reproduces the gauge theory amplitude
of \supermhv\ in the limit when $\a'\to 0$.
To evaluate \stringa, it is convenient to use N=2 notation
and express the vertex operator of \vertU\ as
\eqn\twono{ U_r(z) = \int d\chi_r \int d{ \bar \chi}_r ~\exp
[\pi^a_r\pibar^\ad_r x_{a\ad}(z) +
\chi_r (\pibar_r \psi(z)) + \bar \chi_r (\pibar_r \bar\psi(z))
+ \chi_r\bar \chi_r \e_r^a\pibar_r^\ad \p x_{a\ad}(z) ]}
where $\chi_r$ and $\bar \chi_r$ are Grassmann parameters which are introduced
simply as a technical trick.
Using the free-field OPE's implied by \twoact, one finds
\eqn\corrv{\langle
V(z_1)~ V(z_2)~ V(z_3)~
\prod_{r=4}^N \int dz_r U_r(z_r) \rangle}
$$=
\prod_{r=4}^N \int dz_r d\chi_r d\bar \chi_r
\prod_{r,s} |z_r -z_s|^{\a' p_r p_s}
\exp [\a' {{\pibar_r\pibar_s}\over{z_r-z_s}}
(\chi_r\bar \chi_r (\e_r\pi_s) +\chi_s\bar \chi_s(\e_s\pi_r) + \chi_r \bar \chi_s +\chi_s\bar \chi_r)]$$
where we have chosen a gauge for the $\e_r$'s such that
$\e_r^a \e_{sa}=0$ for all $r$ and $s$.
Note that
\eqn\nodoub{\exp [\a' {{\pibar_r\pibar_s}\over{z_r-z_s}}
(\chi_r\bar \chi_r (\e_r\pi_s) +\chi_s\bar \chi_s(\e_s\pi_r) + \chi_r \bar \chi_s
+\chi_s\bar \chi_r)]}
$$
= 1 +
\a' {{\pibar_r\pibar_s}\over{z_r-z_s}}
(\chi_r\bar \chi_r (\e_r\pi_s) +\chi_s\bar \chi_s(\e_s\pi_r) + \chi_r \bar \chi_s
+\chi_s\bar \chi_r)] $$
and has no double poles when $z_r -z_s\to 0$.

Since each term in the exponential of \nodoub\ is proportional to $\a'$,
these terms can only contribute in the limit $\a'\to 0$
if there appear factors of $1\over\a'$ coming from the integration
over $z_r$. Such factors of $1\over\a'$ can arise from contact terms
when $z_{r-1}\to z_r$ since
\eqn\contact{\int_{z_{r-1}}^{z_{r-1}+\Delta} ~dz_r~
|z_r - z_{r-1}|^{\a' p_r p_{r-1} -1} = (\a' p_r p_{r-1})^{-1}}
for arbitrarily small $\Delta$. So the terms in
\nodoub\ can only contribute if they are proportional to $(z_r-z_{r-1})^{-1}$,
i.e. if they involve neighboring vertex operators.

After integrating over $\prod_{r=4}^N d\chi_r d\bar \chi_r$ and taking
the limit $\a'\to 0$,
one finds that \corrv\ is equal to
\eqn\expone{\lim_{\a'\to 0}\prod_{r=4}^N \int dz_r
|z_r -z_{r-1}|^{\a' p_r p_{r-1}} |z_1 - z_N|^{\a' p_1 p_N}}
$$
\sum_{s=0}^{N-3} ~[~\prod_{t=4}^{N-s}
{{\a'(\pibar_t\pibar_{t-1})(\e_t\pi_{t-1})}\over{z_t-z_{t-1}}}
\prod_{t=N-s+1}^{N}
{{\a'(\pibar_t\pibar_{t+1})(\e_t\pi_{t+1})}\over{z_t-z_{t+1}}} ~] $$
$$=
\sum_{s=0}^{N-3}~[~ \prod_{t=4}^{N-s}
{{(\pibar_t\pibar_{t-1})(\e_t\pi_{t-1})}\over{p_t p_{t-1}}}
\prod_{t=N-s+1}^{N}
{{(\pibar_t\pibar_{t+1})(\e_t\pi_{t+1})}\over{- p_t p_{t+1}}} ~]
$$
$$=
\sum_{s=0}^{N-3}~[~ \prod_{t=4}^{N-s}
{{(\e_t\pi_{t-1})}\over{(\pi_t \pi_{t-1})}}
\prod_{t=N-s+1}^{N}
{{(\e_t\pi_{t+1})}\over{(\pi_{t+1}\pi_t)}} ~]
$$
\eqn\eqfin{= (\pi_3 \pi_1) \prod_{r=3}^N (\pi_r \pi_{r+1})^{-1}.}
Finally, multiplying \eqfin\ by the first line of \stringa,
one reproduces the MHV gauge theory amplitude of \supermhv.

\subsec{Comparison with four-point and five-point gluon amplitudes}

A second check of the conjecture of \stringa\
is that it correctly reproduces the
four-point and five-point gluon scattering when all polarizations
and momenta are four-dimensional.

For four-point scattering, the correlation function in \stringa\
contributes
\eqn\fourcon{\int_{z_3}^{z_1}
dz_4 \sum_{s=1}^3 {{(\ve_4\pi_s)(\pibar_4\pibar_s)}
\over{z_4 - z_s} } \prod_{r,s}
|z_r - z_s|^{\a' (\pi_r\pi_s)(\pibar_r\pibar_s)} }
$$= \int_0^1 dz_4 {{(\pi_3\pi_1)(\pibar_4\pibar_1)}\over
{(z_4 - 1)(\pi_3\pi_4)}}
\prod_{r,s} |z_r - z_s|^{\a' (\pi_r\pi_s)(\pibar_r\pibar_s)} $$
where $\ve_4^a$ has been gauged to
$\ve_4^a = \pi_3^a (\pi_3\pi_4)^{-1}$ and
$(z_1,z_2,z_3)$ have been set to $(1,\infty,0)$.
Multiplying by the first line of \stringa, one obtains the
amplitude
\eqn\multamp{ \eqalign{\widetilde{ \cal A} =&
{{(\pibar_1\pibar_4)}\over{
(\pi_3\pi_4)(\pi_2\pi_3)(\pi_1\pi_2)}}
{{\G(-\a' s +1)\G(-\a' t)}\over{\G(\a' u+1)}}\cr
 \widetilde{ \cal A}=&
\prod_{r=1}^{N=4} (\pi_r\pi_{r+1})^{-1}
{{\G(-\a' s +1)\G(-\a' t+1)}\over{\G(\a' u+1)}}}}
which is the correct open superstring four-point amplitude.

For five-point scattering,
the correlation function in \stringa\
contributes
\eqn\fivecon{\int_{z_3}^{z_1} dz_4 \int_{z_4}^{z_1} dz_5
\prod_{r,s} |z_r - z_s|^{\a' (\pi_r\pi_s)(\pibar_r\pibar_s)} }
$$[\sum_{r\neq 4}
{{(\ve_4\pi_s)(\pibar_4\pibar_s)}
\over{z_4 - z_s} }
\sum_{s\neq 5}
{{(\ve_5\pi_s)(\pibar_5\pibar_s)}
\over{z_5 - z_s} } - {{(\pibar_4\pibar_5)^2}\over{(z_4-z_5)^2}}
-{{(\ve_4\ve_5)(\pibar_4\pibar_5)}\over{\a' (z_4-z_5)^2}}]$$
$$=\int_{1}^{\infty} dz_4 \int_{z_4}^{\infty} dz_5
\prod_{r,s} |z_r - z_s|^{\a' (\pi_r\pi_s)(\pibar_r\pibar_s)} $$
$$[
({{(\pibar_4\pibar_5)(\pi_3\pi_5)}\over{(\pi_3\pi_4)(z_4-z_5)}} +
{{(\pibar_4\pibar_2)(\pi_3\pi_2)}\over{(\pi_3\pi_4)(z_4-z_2)}})
({{(\pibar_5\pibar_4)(\pi_3\pi_4)}\over{(\pi_3\pi_5)(z_5-z_4)}} +
{{(\pibar_5\pibar_2)(\pi_3\pi_2)}\over{(\pi_3\pi_5)(z_5-z_2)}})
- {{(\pibar_5\pibar_4)^2}\over{(z_4-z_5)^2}}]$$
$$=\int_{1}^{\infty} dz_4 \int_{z_4}^{\infty} dz_5
\prod_{r,s} |z_r - z_s|^{\a' (\pi_r\pi_s)(\pibar_r\pibar_s)} $$
$${{(\pi_3\pi_2)}\over{(\pi_3\pi_4)(\pi_3\pi_5)}}
[{{(\pibar_4\pibar_5)(\pibar_5\pibar_2)(\pi_3\pi_5)}\over
{(z_4-z_5)(z_5-z_2)}}
+
{{(\pibar_4\pibar_2)(\pibar_5\pibar_4)(\pi_3\pi_4)}\over
{(z_4-z_2)(z_5-z_4)}}
+
{{(\pibar_4\pibar_2)(\pibar_5\pibar_2)(\pi_3\pi_2)}\over
{(z_4-z_2)(z_5-z_2)}} ]$$
where
$\ve_4^a$ and $\ve_5^a$ have been gauged to
$\ve_4^a = \pi_3^a (\pi_3\pi_4)^{-1}$ and
$\ve_5^a = \pi_3^a (\pi_3\pi_5)^{-1}$, and
$(z_1,z_2,z_3)$ have been set to $(-\infty,0,1)$.

Defining $z_4=x^{-1}$ and $z_5 = (xy)^{-1}$ as in \stieberger,
the integral
\eqn\stie{\int_{1}^{\infty} dz_4 \int_{z_4}^{\infty} dz_5
\prod_{r,s} |z_r - z_s|^{\a' s_{rs} } }
$$
[{A\over{(z_4-z_5)(z_5-z_2)}} + {B\over{(z_4-z_2)(z_5-z_4)}}
+ {C\over{(z_4-z_2)(z_5-z_2)}}] $$
$$= A {{s_{35} f_2 - s_{15} f_1}\over{s_{45}}}
+ B (f_1 -{ {s_{35} f_2 - s_{15}f_1 }\over{s_{45}}})
+C f_1 $$
where $s_{rs} = (\pi_r\pi_s)(\pibar_r\pibar_s)$,
\eqn\fff{f_1 = \int_0^1 dx\int_0^1 dy x^{-1} y^{-1} {\cal I}(x,y),\quad
f_2 = \int_0^1 dx\int_0^1 dy (1-xy)^{-1} {\cal I}(x,y),}
$${\cal I}(x,y) = x^{\a' s_{23}} y^{\a' s_{51}} (1-x)^{\a' s_{34}}
(1-y)^{\a' s_{45}}
(1-xy)^{\a' (s_{12}-s_{34}-s_{45})}.$$

Plugging in
\eqn\vala{A =
{{(\pi_3\pi_2) (\pibar_4\pibar_5)(\pibar_5\pibar_2)}
\over{(\pi_3\pi_4)}},\quad B =
{{(\pi_3\pi_2)(\pibar_4\pibar_2)(\pibar_5\pibar_4)}\over
{(\pi_3\pi_5)}},\quad C =
{{(\pibar_4\pibar_2)(\pibar_5\pibar_2)(\pi_3\pi_2)^2}\over
{(\pi_3\pi_4)(\pi_3\pi_5)}} ,}
using the
identity $\sum_s (\pi_r \pi_s)(\pibar_s\pibar_t)=0$ which follows from the
momentum conservation of $\sum_s p_s=0$,
and multiplying by the factor in the first line of \stringa,
one obtains
\eqn\finalanswer{
\widetilde{ \cal A} ={1 \over
\prod_{s=1}^{N=5}  (\pi_s \pi_{s+1}) } [s_{51}s_{23} f_1 +
(\pi_5\pi_1)(\pibar_1 \pibar_2)(\pi_2 \pi_3)(\pibar_3\pibar_5) f_2] }
which agrees with the
five-point gluon amplitude of \stieberger.

\subsec{BRST operator}

Since the form of the unintegrated operators $V$ and integrated operators
$U$ look very different in \stringa,
it is far from obvious that the superstring formula of
\stringa\ is invariant under cyclic permutations of the $N$
states. In the following subsections, we will give
an argument for this cyclic symmetry
which involves picture-changing operators.
However, these arguments are not rigorous and it would certainly
be useful to better understand this point.

To argue that the prescription has cyclic symmetry, it is convenient
to first define the nilpotent operator
\eqn\brst{Q = \int dz (\l^\a \bar\psi^\ad\p x_{\a\ad}
 + e \psibar_\ad \psibar^\ad + f \l^\a \p\l_\a)}
where $\l^\a$ is a bosonic spinor of conformal weight
$(-\half,0)$ and $e$ and $f$ are conjugate fermions of conformal weight
$(0,0)$ and $(1,0)$ which satisfy the OPE $e(y) f(z)\to \a'(y-z)^{-1}$.
This nilpotent operator will be called a BRST operator for
reasons that will become clear shortly.

Using the free-field OPE's of $(x^{a\ad},\psi_\ad,\psibar_\ad)$,
one can verify that
$Q U_r = \p S_r$ where
\eqn\defV{S_r = (\l\ve_r)(\pibar_r \psibar) e^{i\pi_r\pibar_r x}}
satisfies $QS_r=0$. Furthermore, under $\d\ve^a_r = c\pi_r^a$,
$\d S_r = Q\Omega_r$ where $\Omega_r =
e^{i\pi_r\pibar_r x}$.

Naively, one would compute BRST-invariant tree amplitudes
by evaluating the correlation function of 3 unintegrated vertex
operators $S_r$ and $N-3$ integrated vertex operators $\int dz U_r$.
However, this would give an inconsistent result for two reasons.
Firstly, the 3 unintegrated vertex operators
would contribute three factors of $\psibar$, whereas $\psibar^\ad$
has no zero modes since it has conformal weight $(\half,0)$.
And secondly,
the $-\half$ conformal weight of $\l^\a$ implies that it has
bosonic zero modes on a disk. As will be explained below, $\l^\a$
has 3 bosonic zero modes on a disk and integration over these
non-compact bosonic zero modes would give a factor of
$(\infty)^3$ if the correlation function were defined
using the above vertex operators.

To obtain the appropriate zero mode factors, one needs to
replace the vertex operators $S_r$ of \defV\ with
vertex operators in a lower ``picture''. These
picture-lowered vertex operators $W_r$ will be defined as
\eqn\defW{W_r = (\l\ve_r) \d(\l\pi_r) e^{i\pi_r\pibar_r x}}
where $\d(\l\pi_r)$ denotes a delta-function which constrains
one of the three zero modes of $\l^\a$. It is easy to check
that $QW_r=0$ and that $W_r$ is invariant under the gauge
transformation $\d\ve_r = c\pi_r$.

To understand the relation between $W_r$ of \defW\
and $S_r$ of \defV, note that $S_r = Q\Sigma_r$ where
\eqn\sigmag{\Sigma_r = {{(\l\ve_r)}\over{(\l\pi_r)}}
e^{i\pi_r\pibar_r x}.}
So if $\Sigma_r$ were a well-defined state, $S_r$ would be
BRST-trivial. This situation has an analog in the RNS formalism
since any BRST-closed state $V_{RNS}$ can be written as
$V_{RNS}=Q_{RNS}\Sigma_{RNS}$
where $\Sigma_{RNS} = c\xi\p\xi e^{-2\phi} V_{RNS}$ and
$(\eta e^\phi,\p\xi e^{-\phi})$
is the bosonized version of the $(\gamma,\beta)$ RNS ghosts.
In this case, $\Sigma_{RNS}$ is not a well-defined state since
it depends on the $\xi$ zero mode, i.e. $\eta_0\equiv \int dz \eta$
does not annihilate $\Sigma_{RNS}$. However,
$W_{RNS}=\eta_0\Sigma_{RNS} = c\p\xi e^{-2\phi} V_{RNS}$
is a well-defined state and defines the picture-lowered version of
the vertex
operator. Note that $W_{RNS}=YV_{RNS}$ where
$Y= c\p\xi e^{-2\phi}$ is the picture-lowering
operator satisfying $YX=1$, and $X=\{Q_{RNS}, \xi\}$ is the picture-raising
operator.\fms

To mimic this situation in RNS, suppose that $\l_a\pi_r^a$
is bosonized as
\eqn\bosonz{(\l\pi) = \eta e^\phi.}
This means that
$\Sigma_r$ of \sigmag\ can be expressed as
\eqn\bosons{\Sigma_r =
\xi e^{-\phi}(\l\ve_r) e^{i\pi_r\pibar_r x}.}
Defining the picture-lowered vertex operator
as $W_r = \eta_0 \Sigma_r = e^{-\phi}(\l\ve_r)
e^{i\pi_r\pibar_r x}$, one obtains the operator of \defW\ if
$e^{-\phi}$ is identified as $\d(\l\pi)$. This identification
is very natural and
is analogous to the identification of $e^{-\phi}=\d(\gamma)$ in
the RNS formalism \verlinde.

\subsec{Cyclic symmetry}

In this subsection, it will be shown that the amplitude of \stringa\
can be expressed as
\eqn\stringnew{ \widetilde{ \cal A}  =
\langle W_1(z_1) W_2(z_2) W_3(z_3)
\int_{z_3}^{z_1} dz_4 U_4(z_4) ...
\int_{z_{N-1}}^{z_1} dz_N U_N(z_N) \rangle
}
where the vertex operators $W_r$ and $U_r$ are defined in
\defW\ and \vertU\ and the correlation function in \stringnew\
includes functional integration over the $\l^\a$ zero modes.
Since the unintegrated vertex operators $W_r$ and the
integrated vertex operators $U_r$ are related by picture-changing
operators, one expects to be able to use the usual picture-changing arguments
of \fms\
to prove that \stringnew\ is invariant
under cyclic symmetry.

Since $\l^a$ has conformal weight $-\half$, each component of
$\l^a$ has two zero modes on a disk, i.e.
$\l^a(z) = A^a +z B^a$ where $A^a$ and $B^a$ are zero modes.
However, the BRST operator and all vertex operators
are invariant under the rescaling
\eqn\rescaling{\l^a \to C\l^a,\quad \psibar^\ad \to C^{-1} \psibar^\ad,
\quad \psi^\ad \to C \psi^\ad,\quad e \to C^2 e,\quad f \to C^{-2} f,}
so one of the four zero modes can be gauged away.
The integral over the remaining three zero modes can be easily
performed using the result that
\eqn\perform{\int dA^1 dA^2 dB^1 \l^a(z_1)\l^b(z_2)\l^c(z_3)
\d(\l(z_1)\pi_1) \d(\l(z_2)\pi_2) \d(\l(z_3)\pi_3)}
$$ =
\pi_1^a \pi_2^b \pi_3^c (\pi_1\pi_2)^{-1}(\pi_2\pi_3)^{-1}(\pi_3\pi_1)^{-1}.$$
After plugging \perform\ into \stringnew, one easily verifies that
\stringnew\ reproduces the amplitude prescription of \stringa. So
assuming that the picture-changing manipulations of \fms\ can be applied
to this situation,
\stringa\ has been shown to be invariant under cyclic permutations
of the vertex operators.

\appendix{B}{ Cusp solution for the brane regularization }

In this appendix we find the classical solution describing a string
ending on  a cusp that is sitting at $z= \epsilon$, near the boundary of
$AdS$ space. This is a generalization of the solution in \KruczenskiFB\
  which
describes the case $\epsilon =0$.

We   focus on an $AdS_3$ subspace of $AdS_5$ which is
 parametrized by $x^\pm $ and the radial coordinate
$z$. We want to find the surface that ends on the cusp given by
$x^+ x^-=0$ (only the part in the forward lightcone) and at $z =
\epsilon$.

We assume boost invariance so that the solution depends only on
one variable.
Let us define variables so that $x^\pm = e^{ \tau   \pm \sigma} $
and $ z = e^\tau w(\tau)$. Then the action is
  \aldayjmtwo
\eqn\acthad{ S = \int d\tau {\sqrt{ ( w' + w )^2 -1 } \over w^2} }
The first integral is given by
 \eqn\consen{ c = { w ( w +
w') -1 \over w^2  \sqrt{ ( w' + w )^2 -1 } } } Solving for $w'$ we get
  \eqn\solvforw{ w' = -{ (w^2 -1 - c^2 w^4) +
c w \sqrt{ 1 -w^2 + c^2 w^4} \over w ( c^2 w^2 -1 ) } }
The usual cusp solution is $w=\sqrt{2}$ and $c=1/2$.
We want a solution where $z = \epsilon$ at $\tau = - \infty$. In
this case \eqn\zsmmal{ w = \epsilon e^{-\tau} + 1 + \cdots }
should be the behavior as $\tau \to -\infty$.  It is possible to see
that one can find a solution which obeys these boundary conditions and
asymptotes to the usual cusp solution for large $\tau$ only for $c=1/2$.
In this case  the equation \solvforw\  simplifies and can be solved as
 \eqn\solwbc{
 { e^{\tau} \over \epsilon} = \left( { w +  \sqrt{
2 }  \over w - \sqrt{2} } \right)^{ 1\over \sqrt{2} } { 1 \over 1
+ w }
}
 We get the solution in an implicit form.  We do see that
as $\tau \to -\infty$, then $w \to + \infty$ and we recover \zsmmal . On the other hand as
$\tau \to \infty$, then $ w \to \sqrt{2} $. The range of $w$ is $
( \sqrt{2}, + \infty )$. $w$ becomes  $ \sqrt{2}$ when $e^\tau \gg
\epsilon$. Thus the solution with boundary conditions at $z =
\epsilon$ differs from the solution with boundary condition at
$z=0$ only for $e^\tau$ of the order or smaller than $\epsilon $.

\listrefs

\bye